\pdfoutput=1

\documentclass[11pt,twoside,a4paper,cmspaper,final,collab]{cms-tdr}
\def\svnVersion{490001P}\def\cmsCernNoTag{CERN-EP-2018-289}\def\cmsCernDate{\today}\def\cmsMessage{Published in Physical Review D as \href{http://dx.doi.org/10.1103/PhysRevD.99.032011}{\doi{10.1103/PhysRevD.99.032011.}}}

\begin{document}\cmsNoteHeader{EXO-18-007}

\hyphenation{had-ron-i-za-tion}
\hyphenation{cal-or-i-me-ter}
\hyphenation{de-vices}
\RCS$HeadURL: svn+ssh://svn.cern.ch/reps/tdr2/papers/EXO-18-007/trunk/EXO-18-007.tex $
\RCS$Id: EXO-18-007.tex 489973 2019-02-23 21:01:02Z jingyu $
\newlength\cmsFigWidth
\newlength\cmsTabSkip\setlength{\cmsTabSkip}{1ex}
\ifthenelse{\boolean{cms@external}}{\providecommand{\cmsLeft}{Upper\xspace}}{\providecommand{\cmsLeft}{Left\xspace}}
\ifthenelse{\boolean{cms@external}}{\providecommand{\cmsRight}{Lower\xspace}}{\providecommand{\cmsRight}{Right\xspace}}
\ifthenelse{\boolean{cms@external}}{\providecommand{\NA}{\ensuremath{\cdots}}}{\providecommand{\NA}{\ensuremath{\text{---}}}}
\ifthenelse{\boolean{cms@external}}{\renewcommand{\CL}{\ensuremath{\text{C.L.}}\xspace}}{}
\ifthenelse{\boolean{cms@external}}{\providecommand{\CLnp}{\ensuremath{\text{C.L}}\xspace}}{\providecommand{\CLnp}{\ensuremath{\text{CL}}\xspace}}
\cmsNoteHeader{EXO-18-007}
\title{Search for long-lived particles decaying into displaced jets in proton-proton collisions at \texorpdfstring{$\sqrt{s}=13\TeV$}{sqrt(s) = 13 TeV}}

\date{\today}

\abstract{
A search for long-lived particles decaying into jets is presented. Data
were collected with the CMS detector at the LHC from proton-proton collisions
at a center-of-mass energy of $13\TeV$ in 2016, corresponding to an integrated
luminosity of $35.9\fbinv$.  The search examines the distinctive topology of displaced tracks and secondary vertices. The selected events are found to be consistent with standard model predictions.
For a simplified model in which long-lived neutral particles are pair produced and decay to two jets, pair production cross sections larger
than $0.2\unit{fb}$ are excluded at $95\%$ confidence level for a long-lived particle mass larger than $1000\GeV$ and proper decay lengths between 3 and $130\mm$.
Several supersymmetry models with gauge-mediated supersymmetry
breaking or $R$-parity violation, where pair-produced long-lived gluinos or top squarks decay to several final-state topologies
containing displaced jets, are also tested. For these models, in the mass ranges above $200\GeV$, gluino masses up to 2300--2400$\GeV$ and top squark masses up to 1350--1600$\GeV$ are excluded for proper decay lengths approximately between 10 and
$100\mm$. These are the most restrictive limits to date on these models.
}

\hypersetup{
pdfauthor={CMS Collaboration},
pdftitle={Search for long-lived particles decaying into displaced jets in proton-proton collisions at sqrt(s)=13 TeV},%
pdfsubject={CMS},
pdfkeywords={CMS, physics, exotica, long-lived}}

\maketitle
\section{Introduction}

A large number of extensions to the standard model (SM) predict the production of long-lived particles at the CERN LHC that can further decay into
final states containing jets.
The theoretical motivations are extremely rich~\cite{Liu:2015bma}; examples include split supersymmetry (SUSY)~\cite{Giudice:2004tc,Hewett:2004nw,ArkaniHamed:2004yi,Gambino:2005eh,Arvanitaki:2012ps,ArkaniHamed:2012gw}, SUSY with weak $R$-parity violation (RPV)~\cite{Fayet:1974pd,Farrar:1978xj,Weinberg:1981wj,Barbier:2004ez}, SUSY with gauge-mediated supersymmetry breaking (GMSB)~\cite{GIUDICE1999419,Meade:2008wd,Buican:2008ws}, ``stealth SUSY"~\cite{Fan:2011yu,Fan:2012jf}, ``Hidden Valley" models~\cite{Strassler:2006im,Strassler:2006ri,Han:2007ae}, baryogenesis triggered by weakly
interacting massive particles (WIMPs)~\cite{Cui:2011ab,Cui:2012jh,Cui:2014twa} and twin Higgs models~\cite{Chacko:2005pe,Curtin:2015fna,Cheng:2015buv}.

In this paper, we search for long-lived particles decaying into jets, with each long-lived particle having a decay vertex displaced from the production vertex by up to $55\cm$ in the transverse plane.
Events used in this analysis were collected with the CMS detector~\cite{Chatrchyan:2008aa} at the LHC from proton-proton ($\Pp\Pp$) collisions at a center-of-mass energy of $13\TeV$ in 2016,
corresponding to an integrated luminosity of 35.9\fbinv.
The analysis examines dijets formed by jets clustered from energy deposits in the calorimeters. For the displaced-jet signal,
the tracks left by charged particles originating from the decay of a long-lived particle will usually exhibit large displacements with respect
to the primary vertex, allowing the reconstruction of a secondary vertex within the associated dijet.
The properties of the secondary vertex can be utilized to discriminate between the long-lived signatures and the SM backgrounds. Although the objects studied here
are dijets, two separate displaced single jets can pass the selection criteria,
even when each displaced vertex contains only one jet. A variety of models predict long-lived particles decaying into displaced jets and we test
several of them, including SUSY models with GMSB or RPV, as will be discussed in detail in Section~\ref{sec: trigger}.

Results of searches for similar long-lived particle signatures at $\sqrt{s}=8$\TeV have been reported by ATLAS~\cite{Aad:2015uaa,Aad:2015rba}, CMS~\cite{CMS:2014wda,Khachatryan:2014mea,PhysRevD.95.012009}, and LHCb~\cite{Aaij:2014nma,Aaij:2016isa}.
The ATLAS Collaboration has reported on a search at $\sqrt{s}=13$\TeV, which includes a missing transverse momentum requirement~\cite{Aaboud:2017iio}. The CMS Collaboration has reported several long-lived particle searches at $\sqrt{s}=13\TeV$; one doesn't utilize secondary vertex information~\cite{Sirunyan:2017jdo}, and another searches for a pair of displaced vertices within the beam pipe~\cite{Sirunyan:2018pwn}. The search presented in this paper
is designed to be sensitive to multiple final-state topologies containing displaced jets, and is therefore
sensitive to a wide range of long-lived particle signatures.

\section{The CMS detector}

The central feature of the CMS apparatus is a superconducting solenoid of 6\unit{m} internal diameter, providing a magnetic field of 3.8\unit{T}. Within the
solenoid volume are a silicon pixel and strip tracker, a lead tungstate crystal electromagnetic calorimeter (ECAL), and a brass and scintillator
hadron calorimeter (HCAL), each composed of a barrel and two endcap detectors. Muons are detected in gas-ionization chambers embedded in the
steel flux-return yoke outside the solenoid.

The silicon tracker measures charged particles in the pseudorapidity range $\abs{\eta}<2.5$. It consists of 1440 silicon pixel and 15\,148 silicon
strip detector modules. For nonisolated particles of $1<\pt<10\GeV$ and $\abs{\eta}<1.4$, the track resolutions are typically 1.5$\%$ in $\pt$, and
25--90 (45--150)$\mum$ in the transverse (longitudinal) impact parameter~\cite{Chatrchyan:2014fea}.

In the region $\abs{\eta}<1.74$, the HCAL cells have widths of $0.087$ in pseudorapidity and $0.087$ in azimuth. In the $\eta$-$\phi$ plane,
and for $\abs{\eta}<1.48$, the HCAL cells map on to $5{\times}5$ arrays of ECAL crystals to form calorimeter towers projecting radially outward from the nominal interaction point. For $\abs{\eta}>1.74$, the coverage of the towers increases progressively to a maximum of $0.174$ in
$\Delta\eta$ and $\Delta\phi$. Within each tower, the energy deposits in ECAL and HCAL cells are summed to define the calorimeter tower
energies, and are subsequently used to provide the energies and directions of hadronic jets.

Events of interest are selected using a two-tiered trigger system \cite{Khachatryan:2016bia}. The first level (L1), composed of custom hardware processors, uses
information from the calorimeters and muon detectors to select events at a rate of around $100\unit{kHz}$ within a time interval of less than 4$\mus$.
The second level, known as the high-level trigger (HLT), consists of a farm of processors running a version of the full event reconstruction
software optimized for fast processing, and reduces the event rate to around $1\unit{kHz}$ before data storage.

A more detailed description of the CMS detector, together with a definition of the coordinate system used and the relevant kinematic variables,
can be found in Ref. \cite{Chatrchyan:2008aa}.

\section{Data sets and simulated samples}\label{sec: trigger}

Data were collected with a dedicated HLT displaced-jet trigger. At the trigger level, jets are reconstructed from the energy deposits in the calorimeter towers, clustered using the anti-$\kt$ algorithm~\cite{Cacciari:2008gp,Cacciari:2011ma} with a distance
parameter of 0.4. In this process, the contribution from each calorimeter tower is assigned a momentum, the absolute value and the direction
of which are given by the energy measured in the tower and the coordinates of the tower. The raw jet energy is obtained from the sum of the
tower energies, and the raw jet momentum from the vector sum of the tower momenta, which results in a nonzero jet mass. The raw jet energies are
then corrected~\cite{Khachatryan:2016kdb} to establish a relative uniform response of the calorimeter in $\eta$ and a calibrated absolute response in transverse momentum $\pt$.

Events may contain multiple primary vertices, corresponding to multiple $\Pp\Pp$ collisions occurring in the same bunch crossing.
The reconstructed vertex with the largest value of summed physics-object $\pt^{2}$ is taken to be
the primary $\Pp\Pp$ interaction vertex, referred to as the leading primary vertex. The physics objects are the ``jets," clustered using the jet
finding algorithm~\cite{Cacciari:2008gp,Cacciari:2011ma} with the tracks assigned to the vertex as inputs, and the
associated missing transverse momentum, taken as the negative vector sum of the $\pt$ of those jets. More details are given in
Section~9.4.1 of Ref.~\cite{CMS-TDR-15-02}.

The displaced-jet trigger requires an $\HT$ larger than 350$\GeV$, where $\HT$ is defined as the scalar sum of the transverse momenta of all jets satisfying $\pt>40\GeV$ and $\abs{\eta}<2.5$ in the event. The trigger also
requires the presence of at least two jets, each of them satisfying the following requirements:
\begin{itemize}
\item $\pt>40\GeV$ and $\abs{\eta}<2.0$;
\item at most two associated prompt tracks, which are tracks having a transverse impact parameter (with respect to the leading primary vertex) smaller than 1.0$\unit{mm}$; and
\item at least one associated displaced track, defined as a track with a transverse impact parameter (with respect to the leading primary vertex) larger than 0.5\mm, and an
impact parameter significance larger than 5.0, where the significance is the ratio of the impact parameter to its uncertainty.
\end{itemize}

The main background of this analysis arises from the SM events comprised uniquely of jets produced through the strong interaction, referred to as quantum chromodynamics (QCD)
multijet events. The QCD multijet sample is simulated with $\MGvATNLO$ 2.2.2~\cite{Alwall:2014hca} at leading order, which is interfaced with
$\PYTHIA$ 8.212~\cite{Sjostrand:2014zea} for parton showering, hadronization, and fragmentation. Jets from the matrix element calculations are
matched to parton shower jets using the MLM algorithm~\cite{Alwall:2007fs}. The CUETP8M1 tune~\cite{Khachatryan:2015pea} is used for modeling the underlying event. For parton
distribution function (PDF) modeling, the NNPDF3.0 PDF set~\cite{Ball:2014uwa} is used.

One of the benchmark signal models is a simplified model, referred to as the jet-jet model, where long-lived scalar neutral particles X are pair-produced through a $2\to2$
scattering process, mediated by an off-shell $\PZ$ boson propagator. Each X particle decays to a quark-antiquark pair, and is assumed to do so with equal branching fractions to $\cPqu$, $\cPqd$, $\cPqs$, $\cPqc$, and $\cPqb$ quark pairs. Although X could decay to top quark pairs, we 
chose a signature with a simple topology, such that the analysis strategy would be sensitive to a variety of models. Simulation shows that 
exclusion of the top quark pair decay mode leads to only small changes in the signal efficiency. The chosen signature has two displaced vertices, each of them the origin of one
displaced jet pair. The samples are produced with different resonance masses ranging from 50 to 3000$\GeV$, and with different proper decay
lengths ranging from $1\mm$ to 10\unit{m}.

Several SUSY models with long-lived particles are considered, where we mainly focus on testing SUSY particles with masses larger than $200\GeV$. The first is
a GMSB SUSY model~\cite{Liu:2015bma}, in which the gluino is long lived and
then decays to a gluon and a gravitino, referred to as the $\sGlu\to \Pg\sGra$ model. The gravitino is assumed to be the lightest supersymmetric particle (LSP) and manifests itself as
missing transverse momentum. The signature is two displaced vertices, each of them the origin of a single displaced jet and missing transverse momentum.
The samples are produced with gluino masses from 800 to 2500$\GeV$, and a proper decay length varying from 1$\mm$ to 10\unit{m}.

The second is an RPV SUSY model~\cite{Csaki:2011ge} with minimum flavor violation, where the gluino is
long lived and decays to a top quark and a top squark, the top squark is assumed to be virtual and decays to a
strange antiquark and a bottom antiquark through the RPV interaction with strength given by the coupling $\lambda^{\prime\prime}_{323}$~\cite{Barbier:2004ez}, effectively resulting in a three-body decay with a ``multijet'' final-state topology. This model is referred to as the $\sGlu\to\cPqt\cPqb\cPqs$ model. The samples
are produced with gluino masses from 1200 to 3000$\GeV$, and a proper decay length varying from 1$\mm$
to 10\unit{m}.

Other signal models considered include an RPV SUSY model~\cite{Graham2012}, in which the long-lived top squark decays to a bottom quark and a charged lepton
via RPV interactions with strengths given by couplings $\Lamp_{331}$, $\lambda^{\prime}_{332}$, and $\lambda^{\prime}_{333}$~\cite{Barbier:2004ez}, assuming the decay rate to each of the three lepton flavors to be equal, referred to as the $\sTop\to\cPqb\ell$ model. The samples are produced with different top squark masses from 200 to 1600$\GeV$, and a proper decay
length varying from 1$\mm$ to 1\unit{m}.

We also consider another SUSY model motivated by dynamical $R$-parity violation (dRPV)~\cite{Csaki:2013jza, Csaki:2015fea}, where the long-lived top
squark decays to two down antiquarks via RPV interaction with strength given by a nonholomorphic RPV coupling $\eta^{\prime\prime}_{311}$~\cite{Csaki:2015uza},
referred to as the $\sTop\to\cPaqd\cPaqd$ model. The samples
are produced with different top squark masses from 800 to 1800$\GeV$, and proper decay length
varying from 1$\mm$ to 10\unit{m}.

All signal samples are produced with $\PYTHIA$ 8.212, and NNPDF2.3QED~\cite{Ball:2013hta} is used for PDF modeling. When a gluino or top squark is long lived, it will have
enough time to form a hadronic state, an $R$-hadron~\cite{Farrar:1978xj,Farrar:1978rk,Fairbairn:2006gg}, which is simulated with $\PYTHIA$. For underlying event modeling the CUETP8M1 tune is utilized.

Both the background and the signal events are processed with a $\GEANTfour$-based~\cite{Agostinelli:2002hh} simulation for detailed CMS detector
response. To take account of the effects of additional $\Pp\Pp$ interactions within the same or nearby bunch crossings (``pileup"), additional minimum bias
events are overlaid on the simulated events to match the pileup distribution observed in the data.

\section{Event reconstruction and preselection}\label{sec: pre-sel}

The offline jet reconstruction and primary vertex selection follow the same procedures applied at the trigger level (as described in Section~\ref{sec: trigger}), except that the full offline information is used.
After the trigger selection, events are selected offline requiring $\HT>400\GeV$; dijet candidates are formed from all possible pairs of jets in the event, where the jets are required to have transverse momenta $\pt>50\GeV$
and pseudorapidity $\abs{\eta}<2.0$. These selection criteria are chosen so that the online $\HT$ and jet $\pt$ requirements in the trigger are fully efficient.
The track candidates used in this analysis are required to have ``high purity'' and to have transverse momenta $\pt>1\GeV$. The ``high-purity'' selection
utilizes track information (including the normalized $\chi^{2}$
of the track fit, the impact parameters, and the hits in different layers) to reduce the fake rate and is optimized separately for each iteration of the track reconstruction,
so that it is efficient for selecting tracks with different displacements. More details of the ``high-purity'' selection can be found in Section~4.4 of Ref.~\cite{Chatrchyan:2014fea}. The $\eta$ and $\phi$ of the track are determined by the direction of the momentum vector at the closest point to the leading primary vertex.
The tracks are then associated with the jets by requiring $\Delta R<0.5$, where $\Delta R=\sqrt{\smash[b]{(\Delta\eta)^2+(\Delta\phi)^2}}$ and $\Delta\eta$ ($\Delta\phi$) is the difference in $\eta$ ($\phi$) between the jet axis and the track direction. If a track
satisfies $\Delta R<0.5$ for more than one jet, it is associated with the jet with smaller $\Delta R$.

To reconstruct secondary vertices, displaced tracks associated with each dijet candidate are selected by requiring transverse impact parameters (with respect to the leading primary vertex) larger than $0.5\mm$ and transverse impact parameter significances larger than 5.
An adaptive vertex fitter algorithm~\cite{Fruhwirth:2007hz} is then used for reconstructing a possible secondary vertex
(containing at least 2 tracks) with the displaced tracks in each dijet.
The adaptive vertex fitter utilizes an annealing algorithm in which the outlier tracks
are down-weighted for each step, and thus exhibits robustness against outlier tracks. Only secondary vertices with a $\chi^{2}$
per degree-of-freedom ($\chi^{2}/\mathrm{n_{dof}}$) of less than 5.0 are selected. Also, the four-momentum of the vertex is reconstructed assuming the
pion mass for all assigned tracks; the invariant mass of the vertex is required to be larger than $4\GeV$, and the
transverse momentum of the vertex is required to be larger than $8\GeV$, in order to suppress long-lived SM mesons and baryons.

Each dijet candidate is required to have one reconstructed secondary vertex satisfying the above selection criteria. Furthermore, we select the track with the second-highest transverse (two-dimensional) impact parameter (IP) significance among the tracks that are assigned to the
secondary vertex (the highest two-dimensional IP significance is usually more sensitive to the tail of impact parameter distribution in the
background process, and is therefore less powerful). For displaced-jet signatures, where tracks tend to be more
displaced, the two-dimensional IP significance of this selected track will be large. If it is smaller than 15, the dijet candidate is rejected. We also compute the ratio between the sum of energy for all the tracks
assigned to the secondary vertex and the sum of the energy for all the tracks associated with the two jets. This ratio is expected to be large for displaced-jet signatures, therefore dijet candidates with a ratio smaller than 0.15 are rejected.

An additional variable, $\zeta$, is defined to characterize the contribution of prompt activity to the jets. For each track associated with a jet, the primary vertex (including the leading primary vertex and the pileup vertices) with the minimum three-dimensional impact parameter significance to the track is identified. If this minimum three-dimensional impact parameter significance is smaller than 5, we assign the track to this primary vertex. Then for each jet, we compute the track energy contribution from each primary vertex, and the primary vertex with the largest track energy contribution to the jet is chosen. Finally, we define $\zeta$ as
\begin{linenomath}
\begin{equation}
\zeta=\frac{\sum_{\mathrm{track\in PV_{1}}}E_{\mathrm{track}}^{\mathrm{Jet_{1}}}+\sum_{\mathrm{track\in PV_{2}}}E_{\mathrm{track}}^{\mathrm{Jet_{2}}}}{E_{\mathrm{Jet_{1}}}+E_{\mathrm{Jet_{2}}}},
\end{equation}
\end{linenomath}
which is the charged energy fraction of the dijet associated with the most compatible primary vertices. For displaced-jet signatures, $\zeta$ tends to be small since the jets are not compatible with primary vertices. Dijet candidates with $\zeta$ larger than 0.2 are rejected.

We do not require the secondary vertex to contain tracks from both jets in the dijet candidate. Two displaced single jets originating from
two separate displaced vertices can be paired together and pass the selection, thus the search can be sensitive to long-lived particles decaying to
a single jet (as in the $\sGlu\to\Glu\sGra$ model).

The preselection criteria of the analysis are summarized in Table~\ref{tab: presel}. The variables used in the preselection are checked in data and QCD multijet MC events, and are found to be well-modeled
in the MC events.

\begin{table*}[tb]
\centering
\topcaption{Summary of the preselection criteria}
\label{tab: presel}
\begin{scotch}{lc}
Secondary-vertex/dijet variable & Requirement \\\hline\noalign{\smallskip}
 Vertex $\chi^{2}/\mathrm{n_{dof}}$ & $<$5.0 \\
 Vertex invariant mass & $>$$4\GeV$ \\
 Vertex transverse momentum & $>$$8\GeV$ \\
Second largest two-dimensional IP significance & $>$15 \\
Vertex track energy fraction in the dijet & $>$0.15 \\
$\zeta$ (charged energy fraction associated with compatible primary vertices)   & $<$0.20 \\\noalign{\smallskip}
\end{scotch}
\end{table*}

\section{Event selection and background prediction}\label{sec: sel}

In addition to the secondary vertex reconstruction based on the adaptive vertex fitter, an auxiliary algorithm is explored. For each displaced
track (as defined in Section~\ref{sec: trigger}) associated with the dijet, an expected decay point consistent with the displaced dijet hypothesis is determined by finding the crossing point
between the track helix and the dijet direction in the transverse plane. The displaced tracks associated with the dijet are then clustered based on the expected transverse decay length with respect to the leading primary vertex
$L_{xy}^{\mathrm{exp}}$ using a hierarchical clustering algorithm~\cite{Johnson1967}, in which two clusters are merged together when the smallest expected
transverse decay length difference between the two clusters is smaller than $15\%$ of the transverse decay length ($L_{xy}$) of the secondary vertex.
When more than one cluster is formed after the final step of the hierarchical clustering, the one closest to the secondary vertex is selected.
The cluster root-mean-square (RMS), which is a relative RMS of individual tracks $L_{xy}^{\mathrm{exp}}$ with respect to the secondary vertex $L_{xy}$, is computed to
provide signal-background discrimination:
\begin{linenomath}
\begin{equation}
\mathrm{RMS}_{\mathrm{cluster}}=\sqrt{\frac{1}{N_{\mathrm{tracks}}}\sum_{i=1}^{N_{\mathrm{tracks}}}\frac{(L_{xy}^{\mathrm{exp}}(i)-L_{xy})^{2}}{L_{xy}^{2}}}.
\end{equation}
\end{linenomath}
We then construct a likelihood discriminant based on three variables:

\begin{itemize}
\item vertex track multiplicity;
\item vertex $L_{xy}$ significance;
\item cluster RMS.
\end{itemize}

\begin{figure*}[htbp]
\centering
\includegraphics[width=0.45\textwidth]{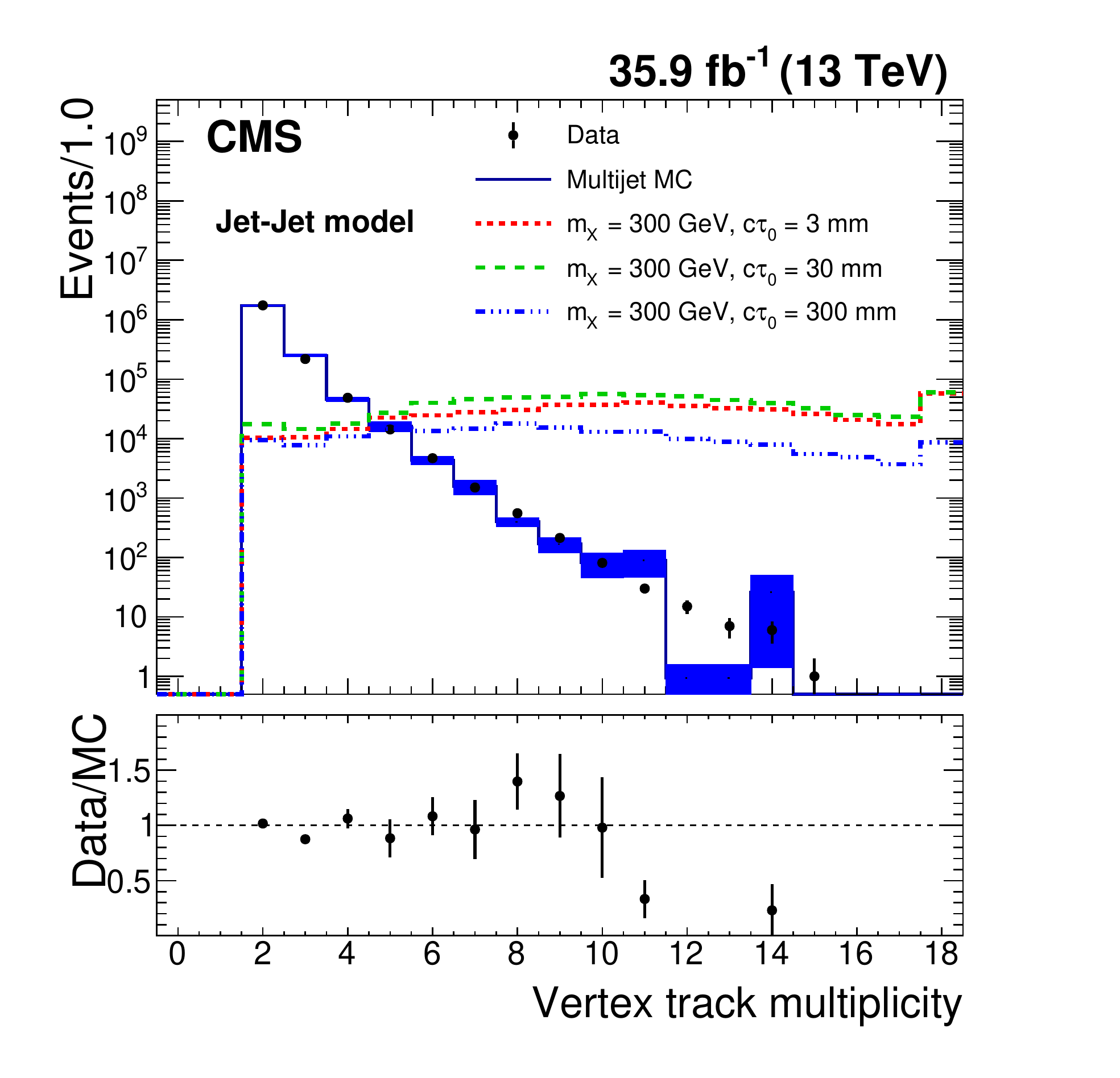}
\includegraphics[width=0.45\textwidth]{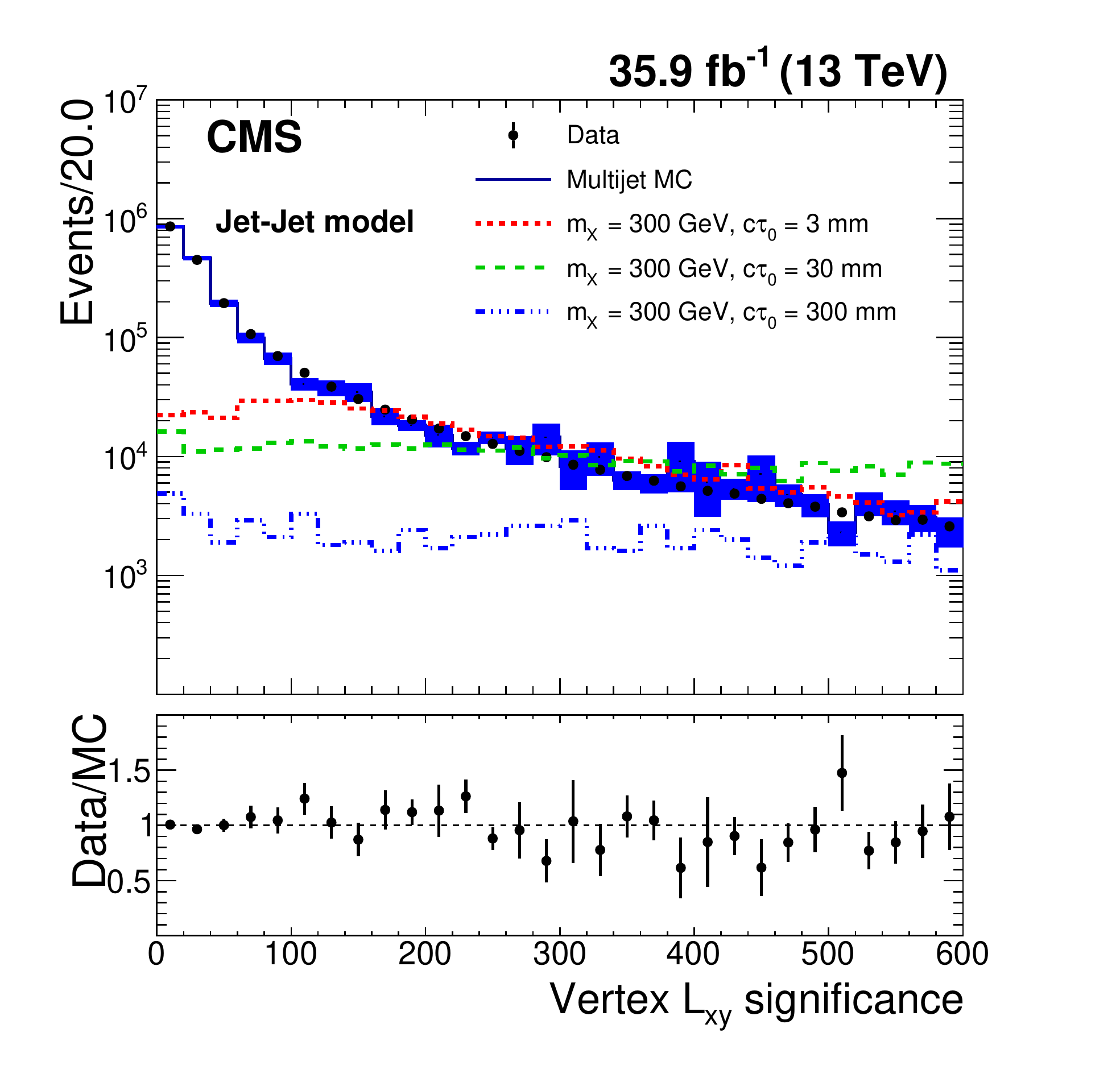}\\
\includegraphics[width=0.45\textwidth]{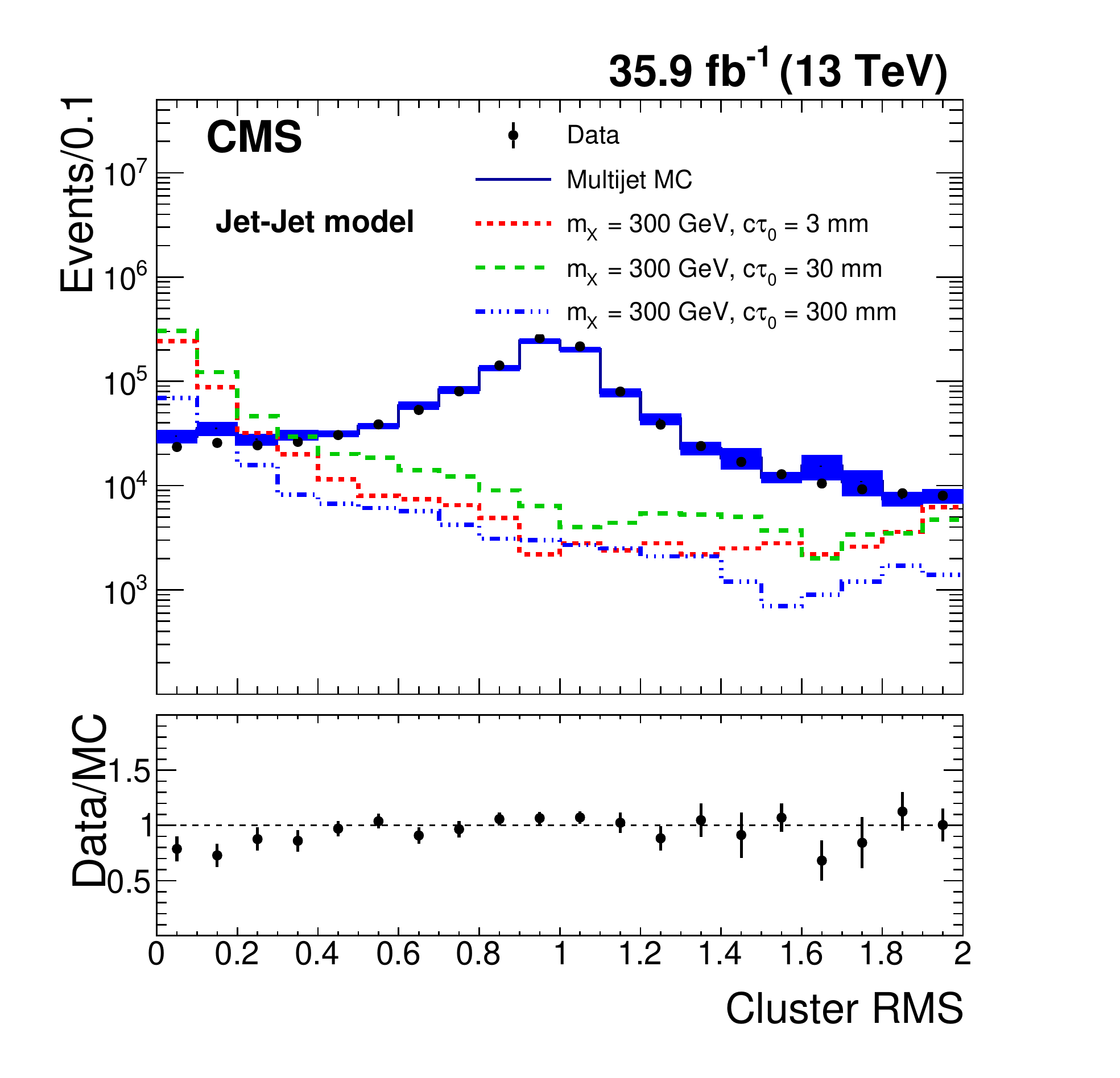}
\includegraphics[width=0.45\textwidth]{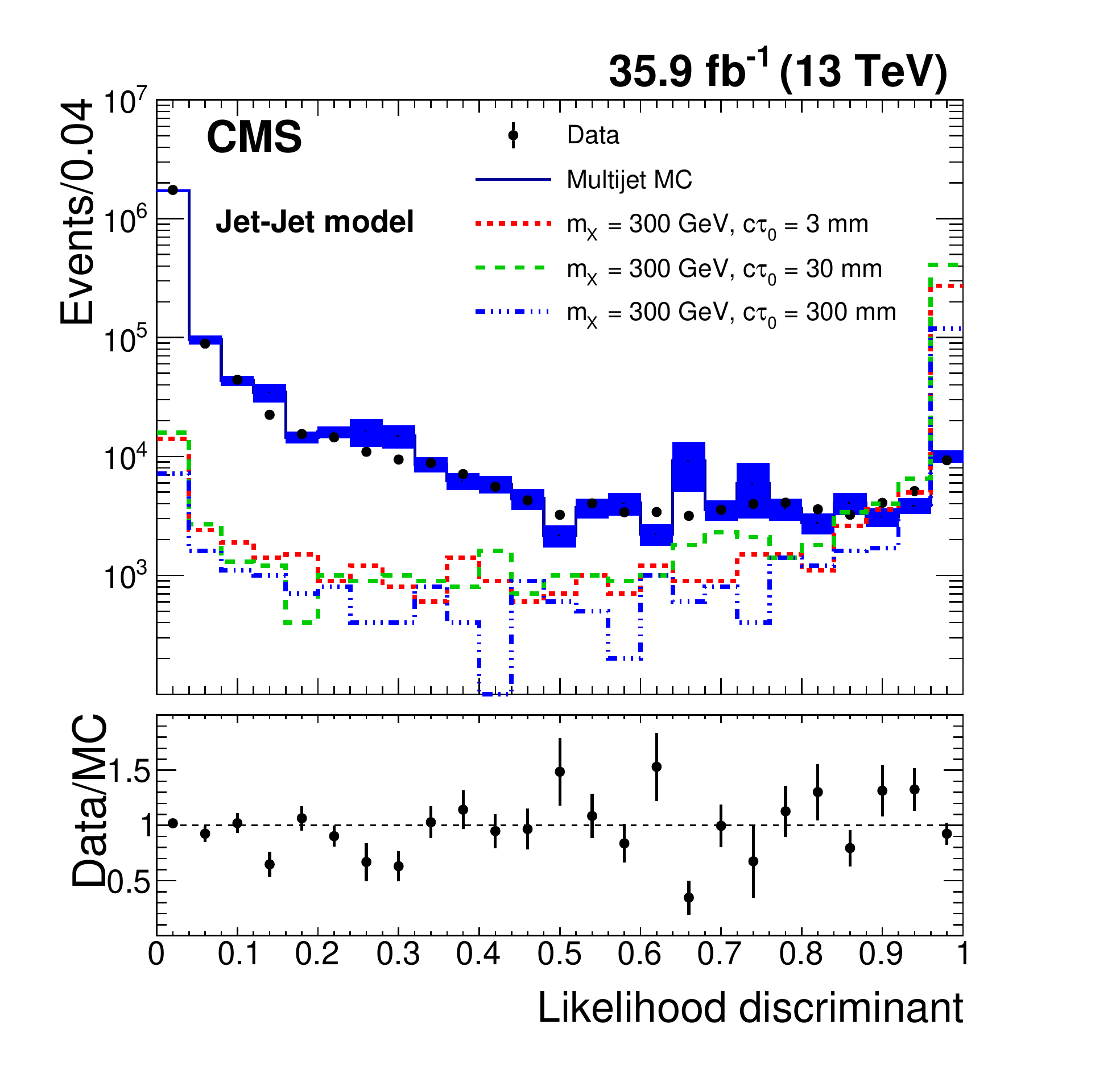}
\caption{
The distributions of vertex track multiplicity (upper left), vertex $L_{\mathrm{xy}}$ significance (upper right), cluster RMS (lower
left), and likelihood discriminant (lower right), for data, simulated QCD multijet events, and simulated signal events. The lower panel of each
plot shows the ratio between the data and the simulated QCD multijet events. Data and simulated events are selected with the displaced-jet trigger.
The offline $\HT$ is required to be larger than 400\GeV, and the jets are required to have $\pt>50\GeV$ and $\abs{\eta}<2.0$. The error bars and
bands represent the statistical uncertainties of each distribution. Three benchmark signal
distributions are shown (dashed lines) for the jet-jet model with $m_{X}=300\GeV$ and varying lifetimes.
For visualization each signal process is given a cross section, $\sigma$, such that $\sigma\ 35.9\fbinv=1\times10^{6}$.}
\label{fig: dijet_var}
\end{figure*}

The three variables are chosen so that the correlations between them are small. The likelihood discriminant, $\mathcal{L}$, is defined as
\begin{linenomath}
\begin{equation}
\mathcal{L}=\frac{p_{\mathrm{S}}}{p_{\mathrm{S}}+p_{\mathrm{B}}}=\frac{1}{1+p_{\mathrm{B}}/p_{\mathrm{S}}}=\frac{1}{1+\prod_{i}p_{\mathrm{Bi}}/p_{\mathrm{Si}}},
\end{equation}
\end{linenomath}
where $p_{\mathrm{S}}$($p_{\mathrm{B}}$) is the probability distribution function of the signal (background), and $i$ is the label for different variables. Simulated jet-jet model events and simulated QCD multijet events are used to derive the probability distribution functions, where
jet-jet model events with $m_{\mathrm{X}}=300$ and $1000\GeV$, and with $c\tau_{0}=1$, 3, 10, 30, 100, 300, and $1000\mm$ are added together to derive $p_{\mathrm{S}}$. When building the likelihood discriminant the trigger requirement is removed, since the number of
simulated events is limited. Figure~\ref{fig: dijet_var} shows the distributions of the three variables used to
build the likelihood discriminant, as well as the discriminant itself, with selections on $\HT$ and jet kinematic variables applied. Simulated signal events for the jet-jet model with $m_{\mathrm{X}}=300\GeV$ at
different proper decay lengths $c\tau_{0}$ are also shown for comparison.

Two other variables are utilized in the event selection. One is the number of three-dimensional prompt tracks in a single jet, where three-dimensional prompt tracks
are the tracks with three-dimensional impact parameters (with respect to the leading primary vertex) smaller than $0.3\mm$. The other is the jet energy fraction carried by two-dimensional prompt tracks, referred to as the
charged prompt energy fraction, where
two-dimensional prompt tracks are those having transverse impact parameters (with respect to the leading primary vertex) smaller than $0.5\mm$.

If more than one dijet candidate passes the preselections described in Section~\ref{sec: pre-sel},
the one with the largest track multiplicity is selected. When the track multiplicities are the same, the one with the
smallest $\chi^{2}$ per degree-of-freedom is selected. The candidate is then required to pass three final selection criteria.
The first makes a selection on the number of three-dimensional prompt tracks and on the charged prompt energy fraction for the leading jet, while the second places a similar requirement on the same variables for the subleading jet. The third makes a selection on the discriminant variable $\mathcal{L}$. The three selection criteria are chosen such that the correlations between them are small for background events. The numerical values of the selection criteria are chosen by optimizing the signal sensitivity for the jet-jet model across different
proper decay lengths (1--1000$\mm$) and different X masses (100--1000$\GeV$). The final selection criteria are determined to be

\begin{itemize}
\item Selection 1: for the leading jet in the dijet candidate, the number of three-dimensional prompt tracks is smaller than 2, the charged prompt energy fraction is smaller than 15\%;
\item Selection 2: for the subleading jet in the dijet candidate, the number of three-dimensional prompt tracks is smaller than 2, the charged prompt energy fraction is smaller than 13\%; and
\item Selection 3: $\mathcal{L}$ is larger than 0.9993.
\end{itemize}

\begin{table}
\centering
\topcaption{The definition of the different regions used in the background estimation.}
\label{tab: abcdefgh}
\begin{scotch}{cccc}
Region &Selection 1 &Selection 2 &Selection 3 \\ \hline
A      &Fail        &Fail        &Fail        \\
B      &Pass        &Fail        &Fail        \\
C      &Fail        &Pass        &Fail        \\
D      &Fail        &Fail        &Pass        \\
E      &Fail        &Pass        &Pass        \\
F      &Pass        &Fail        &Pass        \\
G      &Pass        &Pass        &Fail        \\
H      &Pass        &Pass        &Pass        \\
\end{scotch}
\end{table}

For the jet-jet model, when $m_{\mathrm{X}}=1000\GeV$ and after all the selection criteria are applied, the signal efficiencies for proper decay lengths $c\tau_{0}=1$, 10, 100, and 1000$\mm$ are
9.7, 57, 45, and 7.8$\%$, respectively. When $m_{\mathrm{X}}=100\GeV$, the signal efficiencies for $c\tau_{0}=1$, 10, 100 and, 1000$\mm$ are 0.9, 4.4, 1.6, and 0.2$\%$, respectively. More details of
the signal efficiencies for different signal models can be found in Tables~\ref{tab: eff_JetJet}--\ref{tab: eff_tdd} of Appendix~\ref{app:suppMat}.

Based on the three selections above, eight nonoverlapping regions are defined (regions A--H), as shown in Table~\ref{tab: abcdefgh}. The signal region
(region H) is defined for events passing all three selections. The rest of the regions (A--G) are when events fail one or more of the three selections. The background estimate relies on the three selection criteria having little
correlation between them. The background yield in the signal region H is predicted by different ratios of event counts in regions A--G, where the ratio G(D+E+F)/(A+B+C) uses the fraction of events passing to those failing the likelihood discriminant selection
(selection 3) and is taken as the central value of the predicted background events. Three additional ratios are evaluated using the events failing
one or both of the other two selections (selections 1 and 2):

\begin{itemize}
\item cross-check 1: G(D+E)/(A+C), uses events that fail selection 1;
\item cross-check 2: G(D+F)/(A+B), uses events that fail selection 2; and
\item cross-check 3: G(E+F)/(B+C), uses events that fail either selection 1 or selection 2.
\end{itemize}

These cross-checks provide an important test of the robustness of the background prediction and the assumption that the three selection criteria
are minimally correlated. Differences between the predictions obtained with the nominal method and the cross-checks are also used to estimate the systematic uncertainty in the
background prediction.

The nominal background prediction and the cross-checks are first tested with simulated QCD multijet events, and are found to be robust against different numerical values
for the selection criteria. The method is also checked in data by using a control region defined to be independent to the signal region. This is
achieved by inverting the selection on the vertex track energy fraction in the dijet, requiring this fraction to be less than 0.15. In addition,
in order to improve the statistical precision in the control region, the following two requirements are relaxed relative to the baseline selection:

\begin{itemize}
\item number of three-dimensional prompt tracks smaller than 4; and
\item charged prompt energy fraction smaller than 0.4.
\end{itemize}

\begin{figure*}[htb]
\centering
\includegraphics[width=0.45\textwidth]{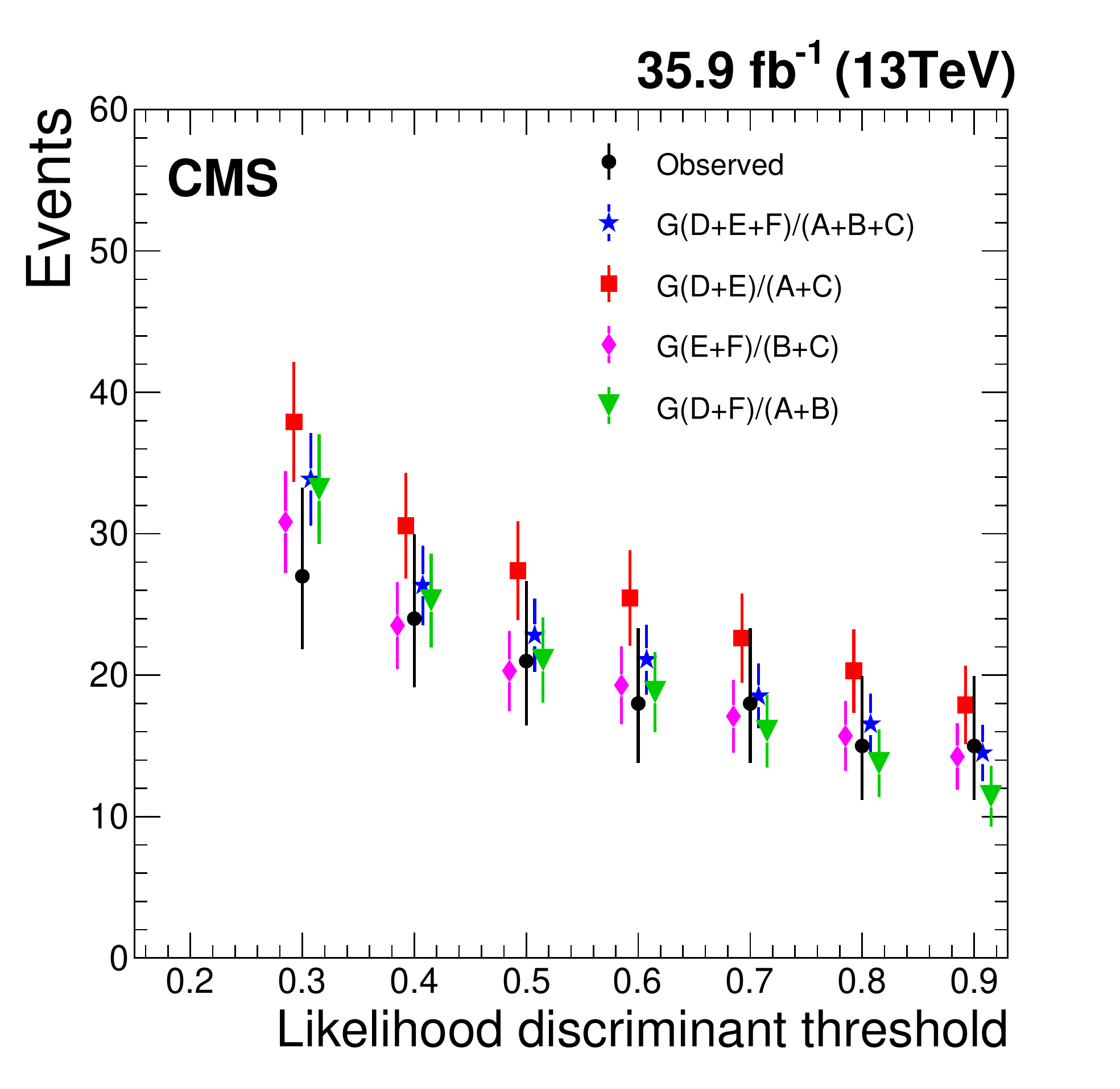}
\includegraphics[width=0.45\textwidth]{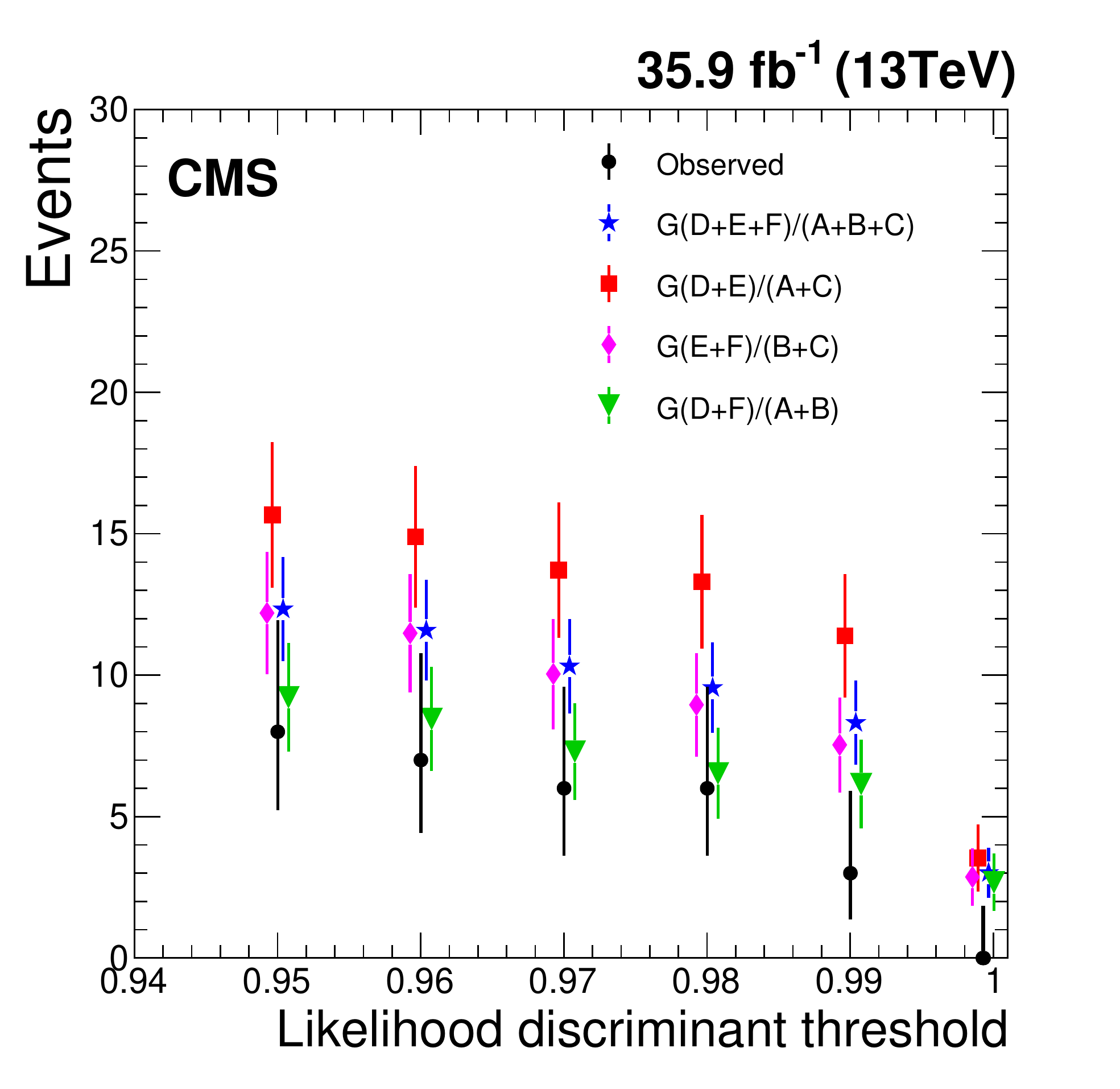}
\caption{Numbers of predicted and observed background events for the nominal background method and the three cross-checks in the control region. Shown are the comparisons for likelihood discriminant thresholds of 0.3, 0.4, 0.5, 0.6, 0.7, 0.8, and 0.9 (left);
and for thresholds of 0.95, 0.96, 0.97, 0.98, 0.99, and 0.9993 (right). The error bars represent the statistical uncertainties of the predictions and the observations. The data points
at different likelihood discriminant thresholds are correlated, since the events passing higher likelihood discriminant thresholds also pass lower likelihood discriminant thresholds.}
\label{fig: data_control_back}
\end{figure*}

The nominal background prediction and cross-checks are then tested in the control region for different threshold values of the likelihood
discriminant. The numbers of predicted and observed background events for the nominal background method and the three cross-checks in the control region are summarized in Fig.~\ref{fig: data_control_back}
and Table~\ref{tab: control_back}. The $p$-value of each observation is computed based on the lower-tail of a Poisson distribution convolved with a
normalized Gaussian function for statistical and systematic uncertainties. The $p$-value is then converted to a $Z$-value using the error function,
\begin{linenomath}
\begin{equation}
Z=\sqrt{2}\erf^{-1}[2p-1],
\end{equation}
\end{linenomath}
which represents the observed significance, expressed as an equivalent number of standard deviations.
The $Z$-values are also listed in the Table~\ref{tab: control_back} for different threshold values of the likelihood discriminant, where the
magnitudes of the $Z$-values are smaller than 1.5 standard deviations.

\begin{table*}[htb]
\centering
\topcaption{The predicted and observed background in the control region for different likelihood discriminant thresholds. The background predictions are shown together with their
statistical (first) and systematic (second) uncertainties (the systematic uncertainties in the background predictions are described in Section~\ref{sec: sys}). The observed significances are also shown in terms of $Z$-values, and are smaller than 1.5 standard deviations.}
\label{tab: control_back}
\begin{scotch}{cccc}
Discriminant threshold & Predicted background & Observed background & $Z$-value \\\hline
0.3  & $33.9\pm3.3\pm4.1$ & 27 & $-0.80$\\
0.5  & $22.8\pm2.6\pm4.6$ & 21 & $-0.14$\\
0.7  & $18.5\pm2.3\pm4.1$ & 18 & $0.05$\\
0.9  & $14.5\pm2.0\pm3.4$ & 15 & $0.24$\\
0.99 & $8.3\pm1.5\pm3.1$ & 3   & $-1.12$\\
0.9993 & $3.0\pm0.9\pm0.5$ & 0 & $-1.40$\\
\end{scotch}
\end{table*}

\section{Systematic uncertainties}\label{sec: sys}

The systematic uncertainties considered include the uncertainty in the background prediction, and the uncertainties in the signal yields. The integrated luminosity uncertainty for the 2016 13\TeV $\Pp\Pp$ collision
data recorded by the CMS detector is determined to be 2.5\%~\cite{CMS-PAS-LUM-17-001}, and is applied as a systematic uncertainty in the signal yields.

The systematic uncertainty in the background prediction is
taken to be the largest deviation from the nominal background prediction (G(D+E+F)/(A+B+C)) to the three cross-checks described in Section~\ref{sec: sel}, and is found to be 11\% for the background yields in the signal region.

The signal efficiencies are calculated with simulated signal samples.
The uncertainty in the efficiency of the online $\HT$ requirement for the trigger emulation is determined by measuring the efficiency
with the events collected with an isolated single-muon trigger. The deviation from full efficiency as a function of
offline $\HT$ for events above the offline $\HT$ threshold is taken as a correction and applied to the signal samples.
Half of each of the corrections for the signal yields are taken as systematic uncertainties, and are calculated
for different masses and proper decay lengths. The largest correction is 5$\%$, thus a systematic
uncertainty of 2.5$\%$ is assigned for all the signal points.

The uncertainty in the efficiency of the online jet $\pt$ requirement is obtained by comparing the per-jet efficiency measured using the data collected with
a prescaled $\HT$ trigger that requires $\HT>325\GeV$, with the efficiency determined from simulated multijet events.
Above the offline $\pt$ threshold, both efficiencies are close to 100$\%$, and the difference between them is
negligible, thus no corresponding systematic uncertainty is assigned.

Similarly, the uncertainty in the efficiency of the  online tracking requirement for the trigger emulation is obtained by comparing the per-jet efficiency measured
using the data collected with the prescaled $\HT$ trigger with the efficiency determined from simulated multijet events. The differences in
the efficiencies between data and simulation are parameterized as functions
of the number of offline prompt and displaced tracks, where the convention of ``prompt'' and ``displaced'' follows
the same definitions described in Section~\ref{sec: trigger}. The difference in the efficiencies is treated as
a bias for the probability of a single jet passing the online tracking requirement, and is applied to the
simulated signal samples. The systematic uncertainty is then determined by computing the variation of the efficiency for
signal events to have at least two jets passing the online tracking requirement. The largest variation is 9$\%$--10$\%$ for the considered signal models in the studied mass-lifetime range, which is taken as the corresponding systematic uncertainty.

To estimate the uncertainty in the offline vertex reconstruction, the events selected with the prescaled $\HT$
trigger are utilized, from which dijet candidates are reconstructed using the same vertex
reconstruction procedure and the same jet kinematics selections as in the offline analysis. We then compare the
data with simulated multijet events in the secondary vertex transverse decay length and
vertex track multiplicity distributions. We find that the main inconsistency between data and multijet simulation lies in the vertex
track multiplicity. A reweighting factor is therefore extracted as a function of the number of tracks in the secondary
vertex, and is interpreted as the correction for the vertex survival probabilities. The correction is then applied
to simulated signal samples vertex-by-vertex, and the systematic uncertainty is obtained by computing the variations of
signal efficiencies after the correction. The uncertainty is found to be 2$\%$--15$\%$ for different signal models in the tested mass-lifetime range.

The uncertainty in the track reconstruction is estimated by studying the track impact parameter measurement in the data and in the multijet
simulation, using the events selected with the prescaled $\HT$ trigger. The possible mismodeling of the impact parameters is taken into account by
varying the impact parameters in the signal samples by the same magnitude. The largest variation in the signal efficiency is taken as the corresponding
uncertainty, and is found to be 14$\%$--20$\%$ for different signal models.

The jet energy scale uncertainty is obtained by varying the jet energy correction~\cite{Khachatryan:2016kdb} by one standard
deviation. The resulting uncertainty is 2$\%$--4$\%$ for the considered signal models.

The uncertainty in the choice of PDF sets is estimated by reweighting the signal events using NNPDF3.0, CT14~\cite{Dulat:2015mca}
and MMHT14~\cite{Harland-Lang:2014zoa} PDF sets, and their associated uncertainty sets~\cite{Butterworth:2015oua,Buckley:2014ana}. The uncertainty in signal efficiencies is quantified by
comparing the efficiencies calculated with alternative PDF sets and the ones with the nominal NNPDF set, and is found to
be 4$\%$--6$\%$ for the considered signal models.

The uncertainty in the selection of the primary vertex is estimated by replacing the leading primary vertex with
the subleading vertex when calculating impact parameters and vertex displacement, where the primary vertices are ordered based on their values of summed physics-object $\pt^{2}$ as described in
Section~\ref{sec: trigger}. The resulting uncertainty in
signal efficiency is found to be 6$\%$--15$\%$ for different signal models in the tested mass-lifetime range.

A summary of different sources of systematic uncertainties in the signal yields
can be found in Table~\ref{tab: sys_unc}. For each signal model, the largest variations due to each source across
the tested mass-lifetime points are taken as the corresponding systematic uncertainties.

\begin{table*}[htb]
\centering
\topcaption{Systematic uncertainties in the signal yields, for each signal model studied. The quoted values reflect the largest variations due to each source for each signal model, in the
studied range of masses and proper decay lengths.}
\label{tab: sys_unc}
\begin{scotch}{lccccc}
\noalign{\smallskip}
Source & Jet-jet model & $\sGlu\to \Pg\sGra$ & $\sGlu\to\cPqt\cPqb\cPqs$ & $\sTop\to \cPqb\ell$ & $\sTop\to\cPaqd\cPaqd$\\\hline
Integrated luminosity                      & 2.5$\%$ & 2.5\% & 2.5\% & 2.5\% &  2.5\% \\
Online $\HT$ requirement        & 2.5$\%$ & 2.5\% & 2.5\% & 2.5\% &  2.5\% \\
Online tracking requirement     & 9\%     & 9\%   &9\%    & 9\%   &  10\% \\
Offline vertexing               & 15\%    & 2\%   & 6\%   & 5\%   &  2\% \\
Track impact parameter modeling & 14\%    & 16\%  & 20\%  & 10\%  &  20\%\\
Jet energy scale                & 4\%     & 2\%   & 2\%   & 2\%   &  2\% \\
PDF                             & 5$\%$   & 6\%   & 4\%   & 5\%   &  4\%\\
Primary vertex selection        & 6$\%$   & 10\%  & 15\%  & 8\%   &  12\%\\[\cmsTabSkip]
Total                           & 24$\%$& 22\%& 28\%& 18\%&  26\%\\
\end{scotch}
\end{table*}

\section{Results}
\subsection{Data in the signal region}

We divide the signal region in bins of $\HT$ and the number of dijets passing
the preselection criteria in order to gain sensitivity to long-lived particles with different masses.
After applying all the selection criteria described in Sections~\ref{sec: pre-sel} and~\ref{sec: sel}, we observe one event in the data,
in accord with the total background prediction of $1.03\pm0.19\stat\pm0.11\syst$ events. This observed event has an $\HT$ of 590$\GeV$;
and yields a secondary vertex candidate, with a transverse decay length of 3.5$\cm$ and a track multiplicity of 10. This is
consistent with the presence of a {\cPqb} quark jet, where the bottom hadron travels in the tracker for an extremely long distance before it decays.

Table~\ref{tab: exp_back} shows the predicted background and observations in the different bins of the signal region, where the sum of the predicted background in the 
four bins is consistent with the total background prediction 
quoted earlier. We find the observed
yield is consistent with the predicted background in all bins, and we use the results in the four bins to set limits on a variety of models.

\begin{table*}[!htb]
\centering
\topcaption{Summary of predicted and observed events in the signal region, for different $\HT$ and number of dijet candidates values.}
\label{tab: exp_back}
\begin{scotch}{cccc}
Selection on $\HT$ & Number of dijets & Expected & Observed\\\hline
$400<\HT<450\GeV$ & 1 & $0.42\pm0.14\stat\pm0.01\syst$ & 0\\
$450<\HT<550\GeV$ & 1 & $0.23\pm0.08\stat\pm0.07\syst$ & 0\\
$\HT>550\GeV$ & 1 & $0.19\pm0.07\stat\pm0.05\syst$ & 1\\
\NA &$>$1 & $0.16\pm0.11\stat\pm0.06\syst$ & 0\\
\end{scotch}
\end{table*}

\subsection{Interpretation of results}

We set upper limits on the production cross section versus mass or lifetime for a given model by computing the 95$\%$ confidence level ($\CL$) associated with each signal point according to the $\CLs$ prescription~\cite{Junk:1999kv,Read:2002hq,Cowan:2010js,CMS-NOTE-2011-005}, using an LHC-style profile likelihood ratio~\cite{Cowan:2010js,CMS-NOTE-2011-005} as the
test statistics. The $\CLs$ values are calculated using the asymptotic approximation~\cite{Cowan:2010js}, and are verified with
full-frequentist results for representative signal points. The signal yields in the four bins in Table~\ref{tab: exp_back} are
utilized to compute the $\CLs$ values, and the systematic uncertainties are taken to be fully correlated across the four bins. The bin where more than one dijet candidate passes the preselection criteria usually brings most of the sensitivity in a given model since it often has the largest signal efficiency.

\begin{figure}[!htb]
\centering
\includegraphics[width=0.49\textwidth]{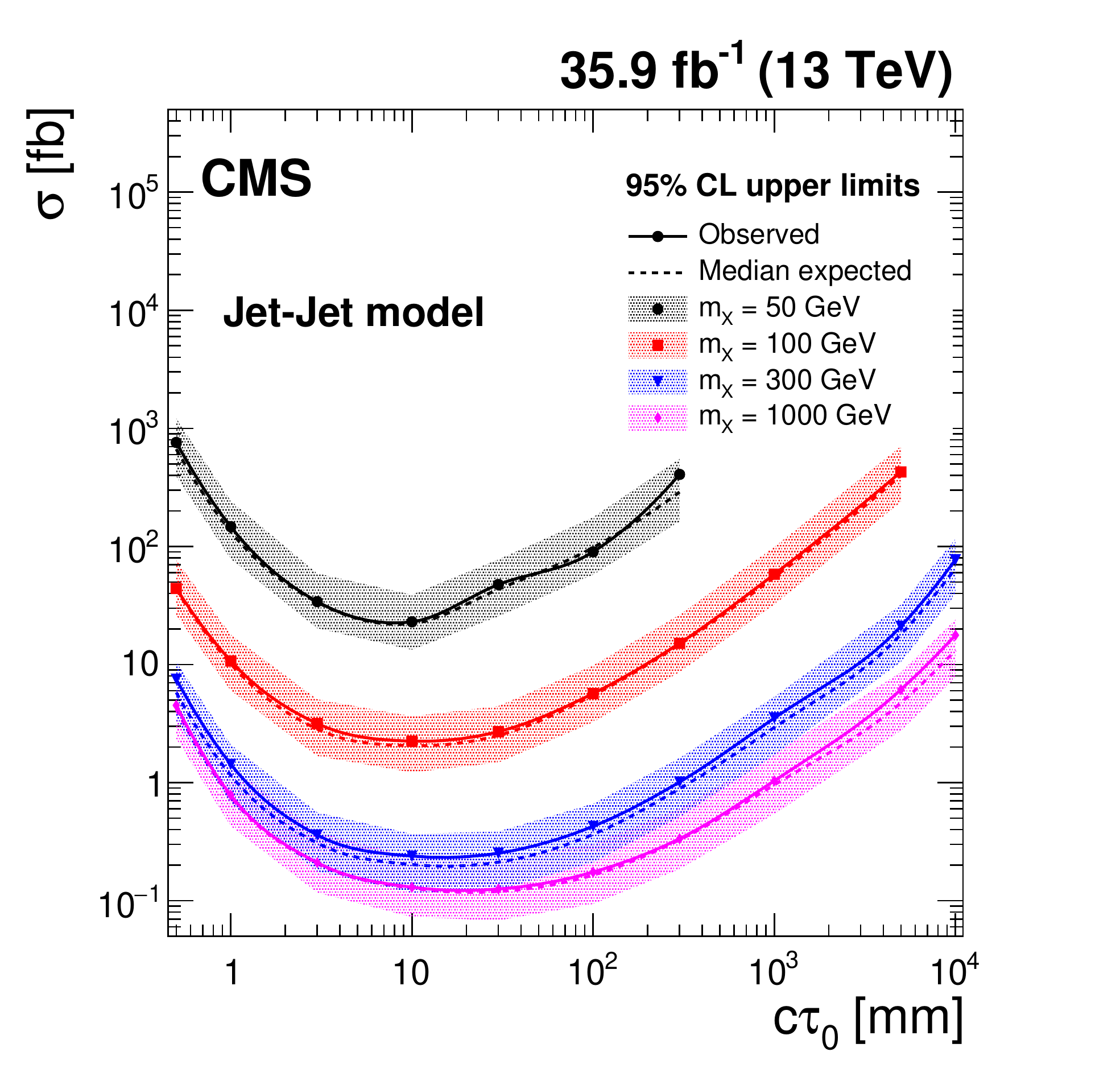}
\caption{The expected and observed 95$\%$ $\CL$ upper limits on the pair production cross section of the long-lived particle X, assuming
a $100\%$ branching fraction for X to decay to a quark-antiquark pair, shown at different particle X masses and
proper decay lengths for the jet-jet model. The solid (dashed) lines represent the observed (median expected) limits. The shaded bands represent the regions
containing $68\%$ of the distributions of the expected limits under the background-only hypothesis. }
\label{fig: expected_limit_XXTo4J}
\end{figure}

Figure~\ref{fig: expected_limit_XXTo4J} presents the expected and observed upper limits (at $95\%$ $\CL$) on the pair production cross section for the jet-jet model
at different scalar particle X masses and proper decay lengths, assuming a $100\%$ branching fraction. The limits are most stringent for $c\tau_{0}$ between 3 and $100\mm$.
For smaller decay lengths, the limits become less restrictive because of the vetoes on prompt activity. Since the tracking efficiency
decreases with larger displacement, the limits also become less stringent for larger decay lengths when $c\tau_{0}>100\mm$.
Pair production cross sections larger than 0.2\unit{fb} are excluded at high mass ($m_{\rm{X}}>1000\GeV$) for proper decay lengths between 3 and 130$\mm$.
The lowest pair production cross section excluded is 0.13\unit{fb}, at $c\tau_{0}=30\mm$ and long-lived particle mass $m_{\rm{X}}>1000\GeV$.

\begin{figure}[tbp]
\centering
\includegraphics[width=0.45\textwidth]{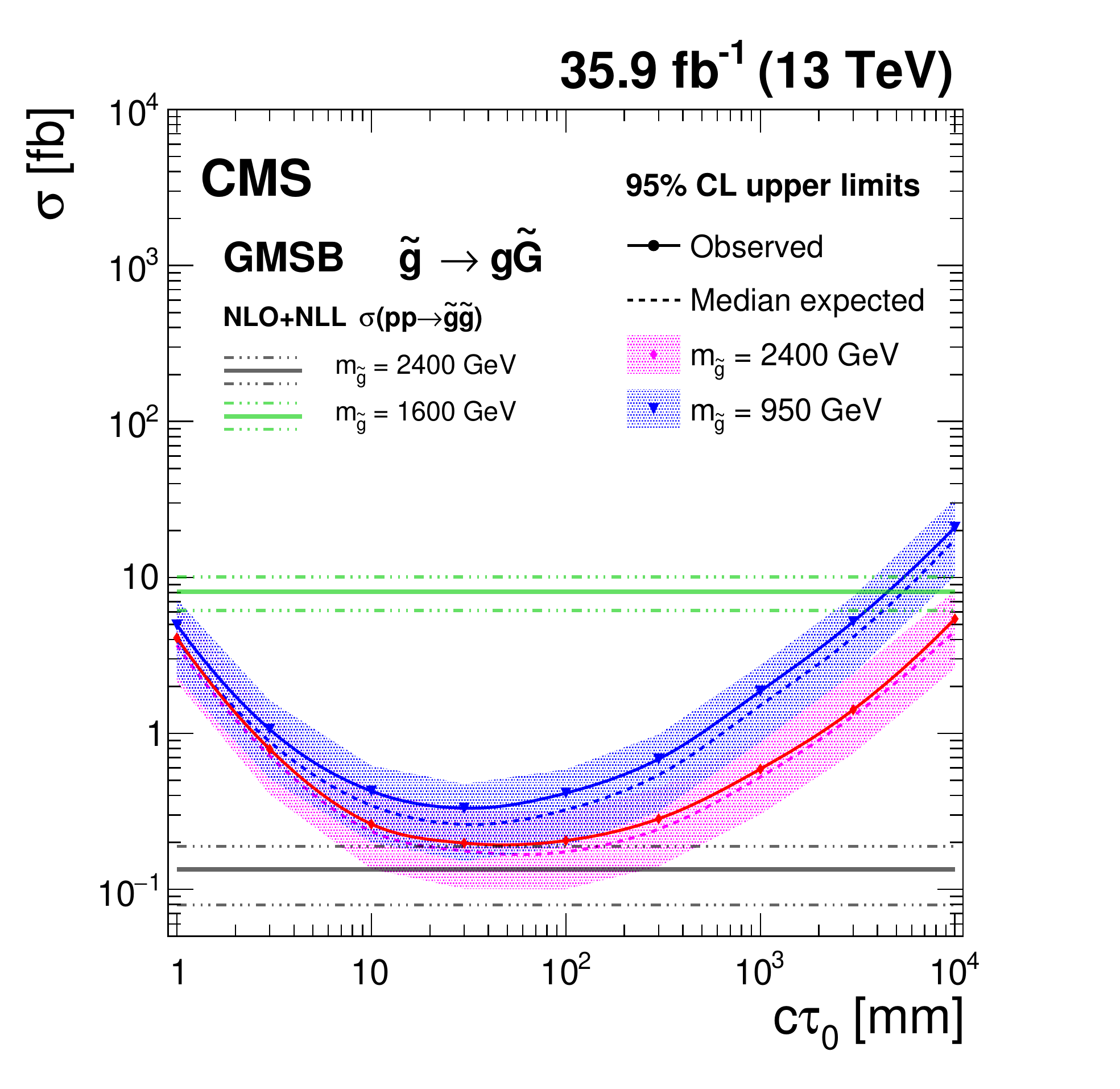}
\includegraphics[width=0.45\textwidth]{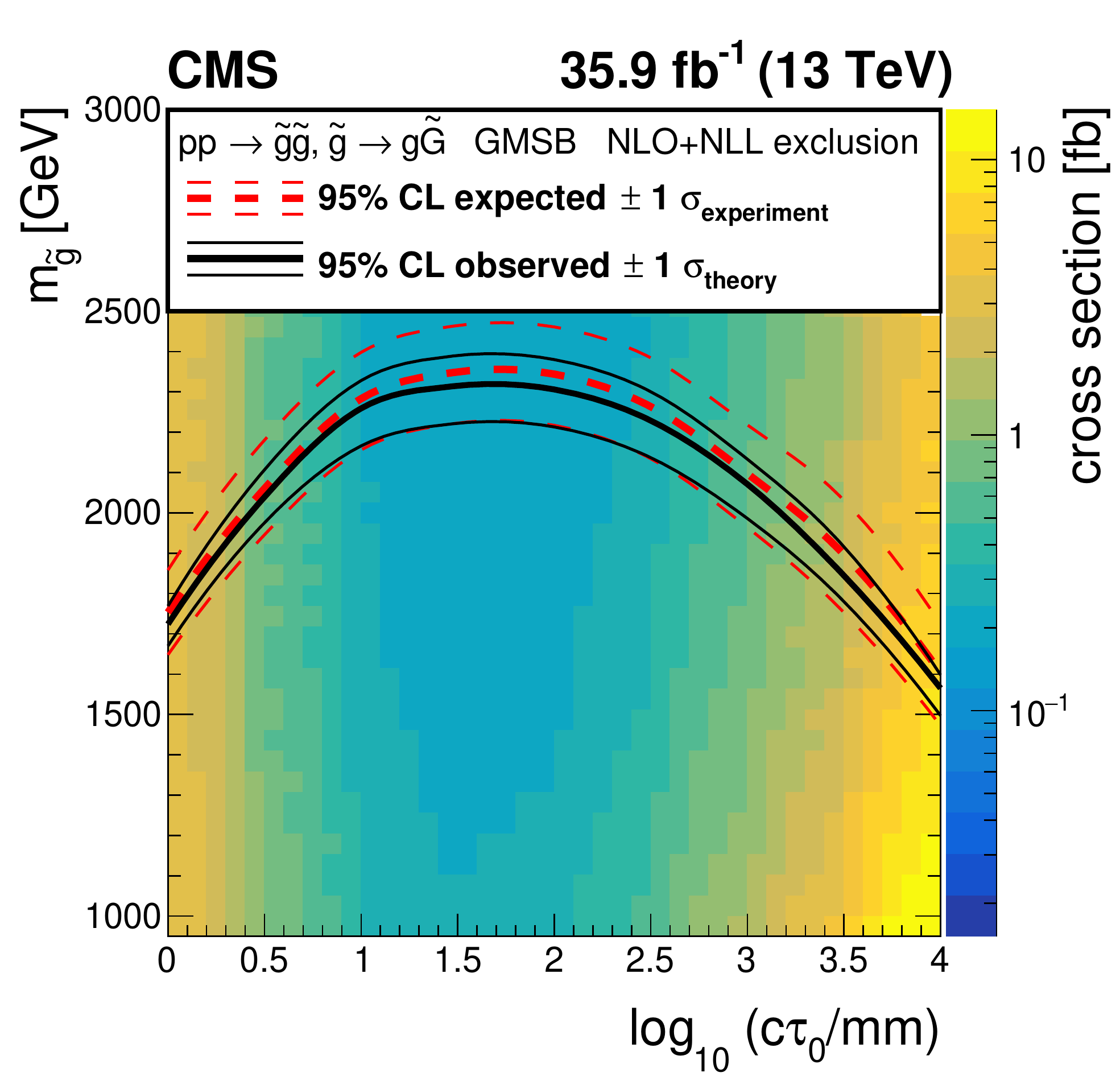}
\caption{\cmsLeft: the expected and observed $95\%$ $\CL$ upper limits on the pair production cross section of the long-lived gluino, assuming a
$100\%$ branching fraction for $\sGlu\to\Glu\sGra$ decays. The horizontal lines indicate the NLO+NLL gluino pair production cross sections for $m_{\sGlu}=2400\GeV$ and $m_{\sGlu}=1600\GeV$, as well as their variations due to the uncertainties in the choices of renormalization scales, factorization scales, and PDF sets. The solid (dashed) lines represent the observed (median expected) limits, the bands show the regions containing $68\%$ of
the distributions of the expected limits under the background-only hypothesis.
\cmsRight: the expected and observed $95\%$ $\CL$ limits for the long-lived gluino model in the mass-lifetime plane, assuming a 100\% branching fraction for $\sGlu\to \Pg\sGra$ decays, based on the NLO+NLL calculation of the gluino pair production cross section at $\sqrt{s}=13\TeV$. The thick solid black
(dashed red) line represents the observed (median expected) limits at 95$\%$ $\CLnp$. The thin black lines represent the change in the observed limit
due to the variation of the signal cross sections within their theoretical uncertainties. The thin red lines indicate the region containing $68\%$ of
the distribution of the expected limits under the background-only hypothesis.}
\label{fig: expected_limit_GMSB}
\end{figure}

Figure~\ref{fig: expected_limit_GMSB} presents the expected and observed upper limits on the pair production cross section of long-lived gluino in
the GMSB $\sGlu\to\Glu\sGra$ model, assuming a $100\%$ branching fraction for the gluino to decay into a gluon and a gravitino.
Although in the $\sGlu\to\Glu\sGra$ signature each displaced vertex is associated with only one jet, the two separate displaced single jets can be paired
together and pass the selections, therefore the analysis is sensitive to this kind of signature. When the gluino mass is
2400$\GeV$, gluino pair production cross sections larger than 0.25\unit{fb} are excluded for proper decay lengths between 10 and 210$\mm$.
When the proper decay length $c\tau_{0}=1\mm$, the upper limit is insensitive to the gluino mass in the tested range since the signal
acceptance is mainly limited by the online prompt track requirement in the displaced-jet trigger. The upper limits on the pair production
cross section are then translated into upper limits on the gluino mass for different proper decay lengths, based on a calculation at the next-to-leading logarithmic accuracy matched to next-to-leading order
predictions (NLO+NLL) of the gluino pair production cross section at $\sqrt{s}=13\TeV$~\cite{Beenakker:1996ch,Kulesza:2008jb,Kulesza:2009kq,Beenakker:2011fu,Borschensky:2014cia} in the limit where all the other SUSY particles are much heavier and decoupled. Gluino masses up to $2300\GeV$
are excluded for proper decay lengths between 20 and $110\mm$. The bounds are the most stringent to date on this model in the tested proper
decay length range.

\begin{figure}[tbp]
\centering
\includegraphics[width=0.45\textwidth]{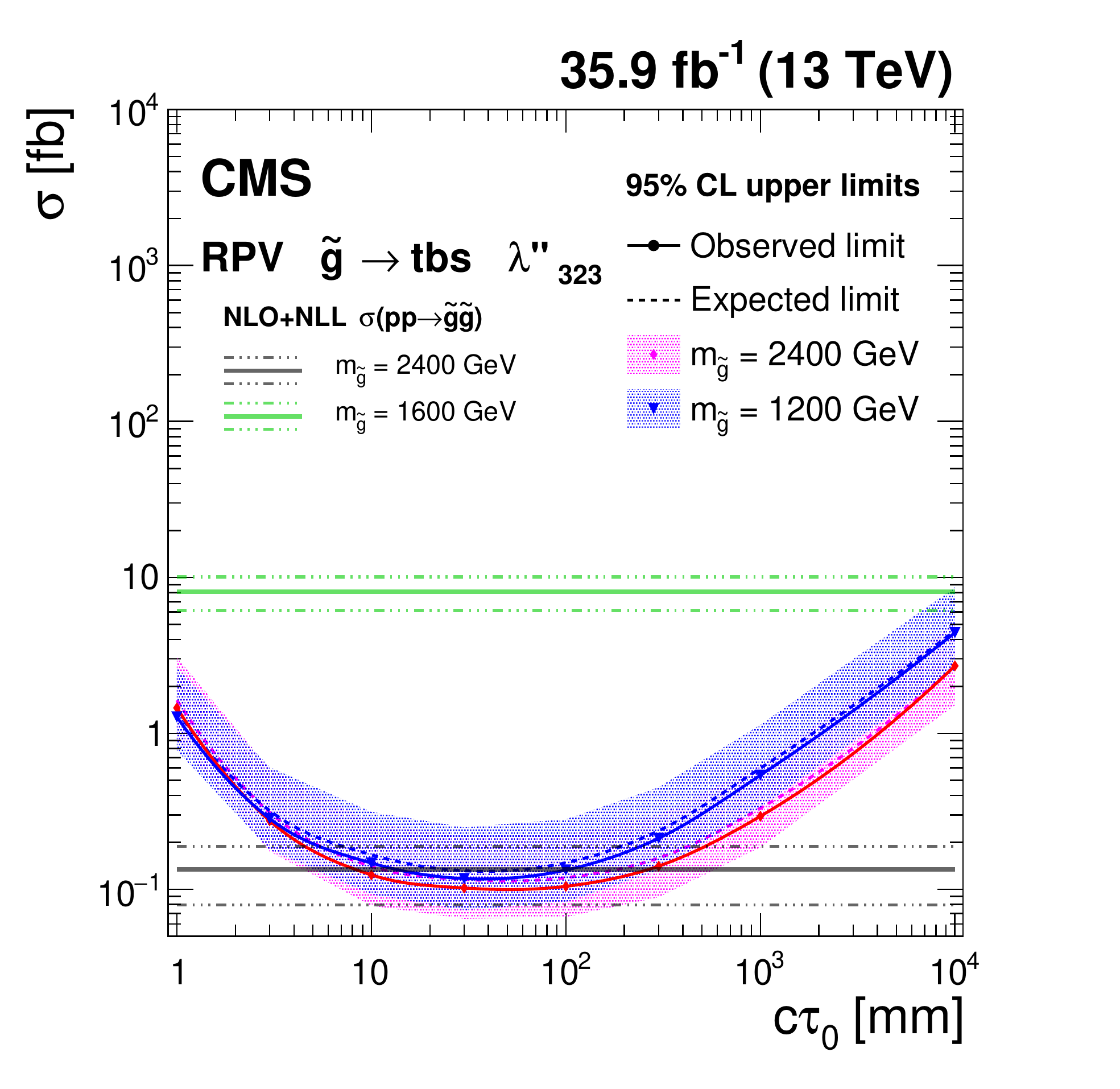}
\includegraphics[width=0.45\textwidth]{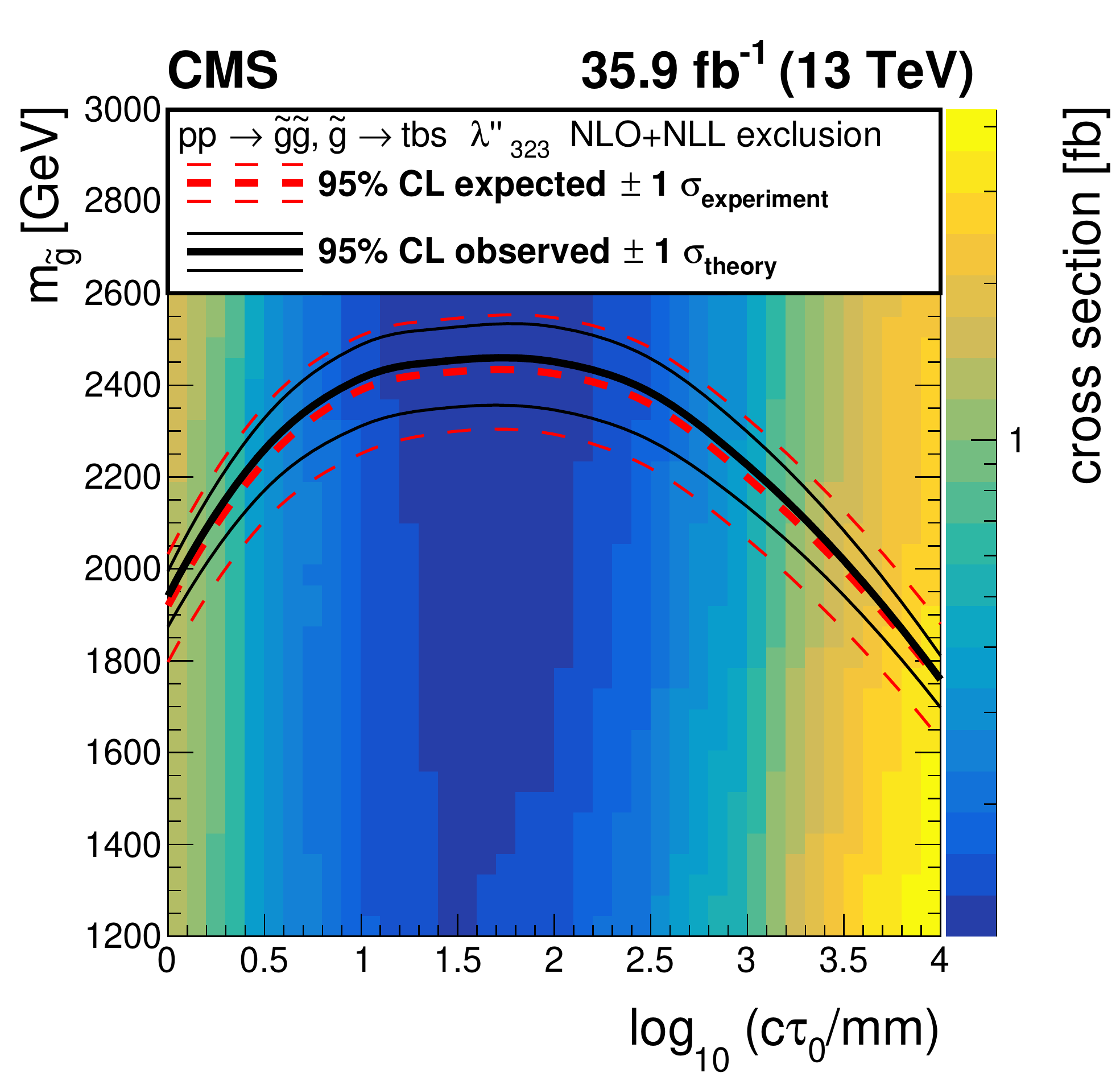}
\caption{\cmsLeft: the expected and observed $95\%$ $\CL$ upper limits on the pair production cross section of the long-lived gluino, assuming a $100\%$ branching fraction for $\sGlu\to\cPqt\cPqb\cPqs$ decays. The horizontal lines indicate the NLO+NLL gluino pair production cross sections for $m_{\sGlu}=2400\GeV$ and $m_{\sGlu}=1600\GeV$, as well as their variations due to the uncertainties in the choices of renormalization scales, factorization scales, and PDF sets. The solid (dashed) lines represent the observed (median expected) limits, the bands show the regions containing $68\%$
of the distributions of the expected limits under the background-only hypothesis.
\cmsRight: the expected and observed $95\%$ $\CL$ limits for the long-lived gluino model in the mass-lifetime plane, assuming a $100\%$ branching fraction
for $\sGlu\to\cPqt\cPqb\cPqs$ decays, based on the NLO+NLL calculation of the gluino pair production cross section at $\sqrt{s}=13\TeV$. The thick solid black (dashed red) line represents the observed (median expected) limits at $95\%$ $\CLnp$. The thin black lines represent the change in the observed limit
due to the variation of the signal cross sections within their theoretical uncertainties. The thin red lines indicate the region containing $68\%$ of
the distributions of the expected limits under the background-only hypothesis.}
\label{fig: expected_limit_GGToNN}
\end{figure}

Figure~\ref{fig: expected_limit_GGToNN} presents the expected and observed
upper limits on the pair production cross section of the long-lived gluino in the RPV $\sGlu\to\cPqt\cPqb\cPqs$ model,
assuming a $100\%$ branching fraction for the gluino
to decay to top, bottom, and strange antiquarks. The upper limits on
the pair production cross section are translated into upper limits on the
gluino mass for different proper decay lengths, based on the NLO+NLL calculation of the gluino pair production cross
section at $\sqrt{s}=13\TeV$~\cite{Beenakker:1996ch,Kulesza:2008jb,Kulesza:2009kq,Beenakker:2011fu,Borschensky:2014cia} in the limit where all the other SUSY particles are much heavier and decoupled. Gluino masses up to $2400\GeV$
are excluded for proper decay lengths between 10 and $250\mm$. The bounds are
currently the most stringent on this model for proper decay lengths between $10\mm$ and 10\unit{m}. A comparison on this model with the
existing CMS search for displaced vertices within the beam pipe~\cite{Sirunyan:2018pwn} can be found in Fig.~\ref{fig: expected_limit_comparison} of Appendix~\ref{app:suppMat}.

\begin{figure}[tbp]
\centering
\includegraphics[width=0.45\textwidth]{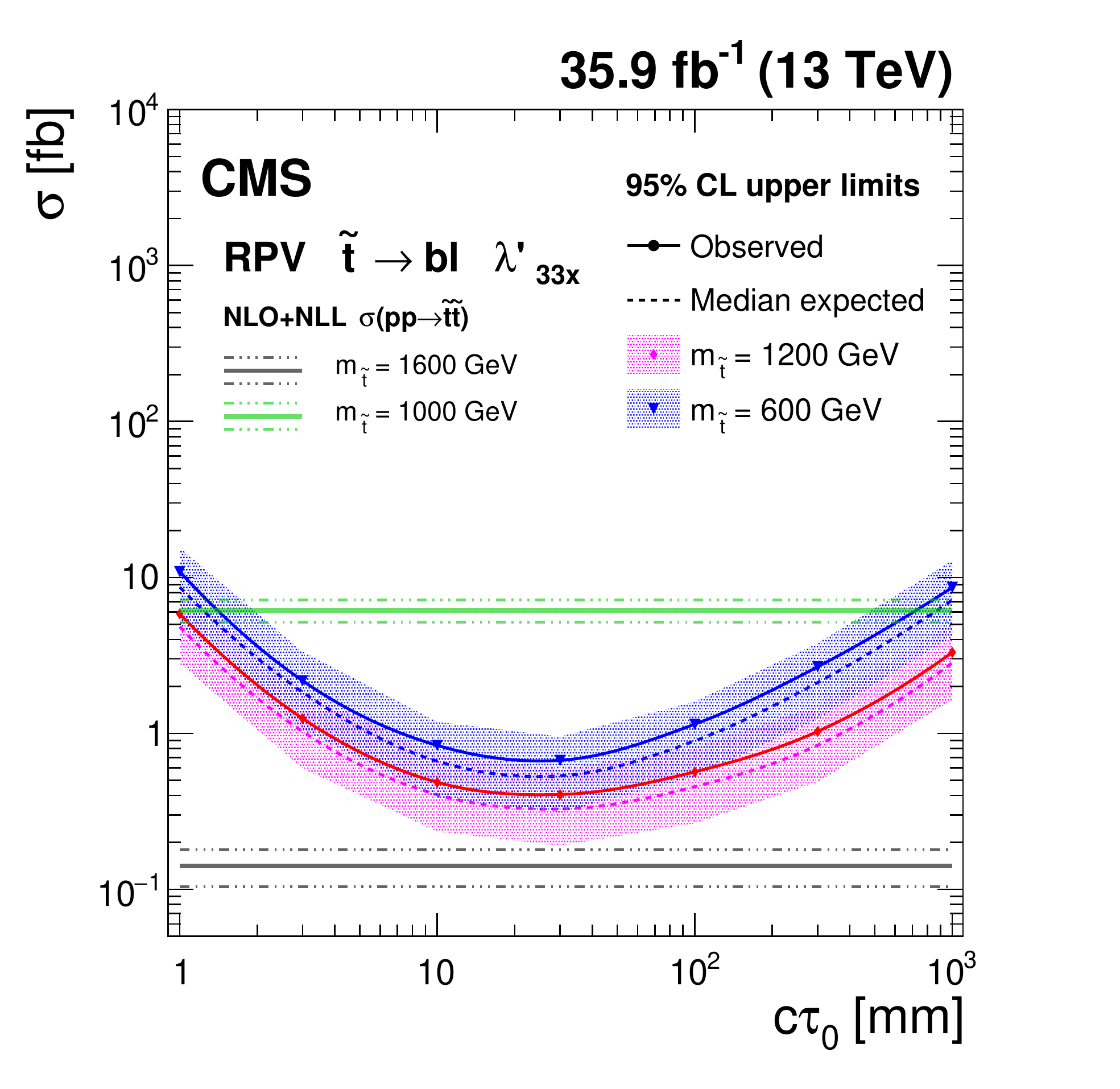}
\includegraphics[width=0.45\textwidth]{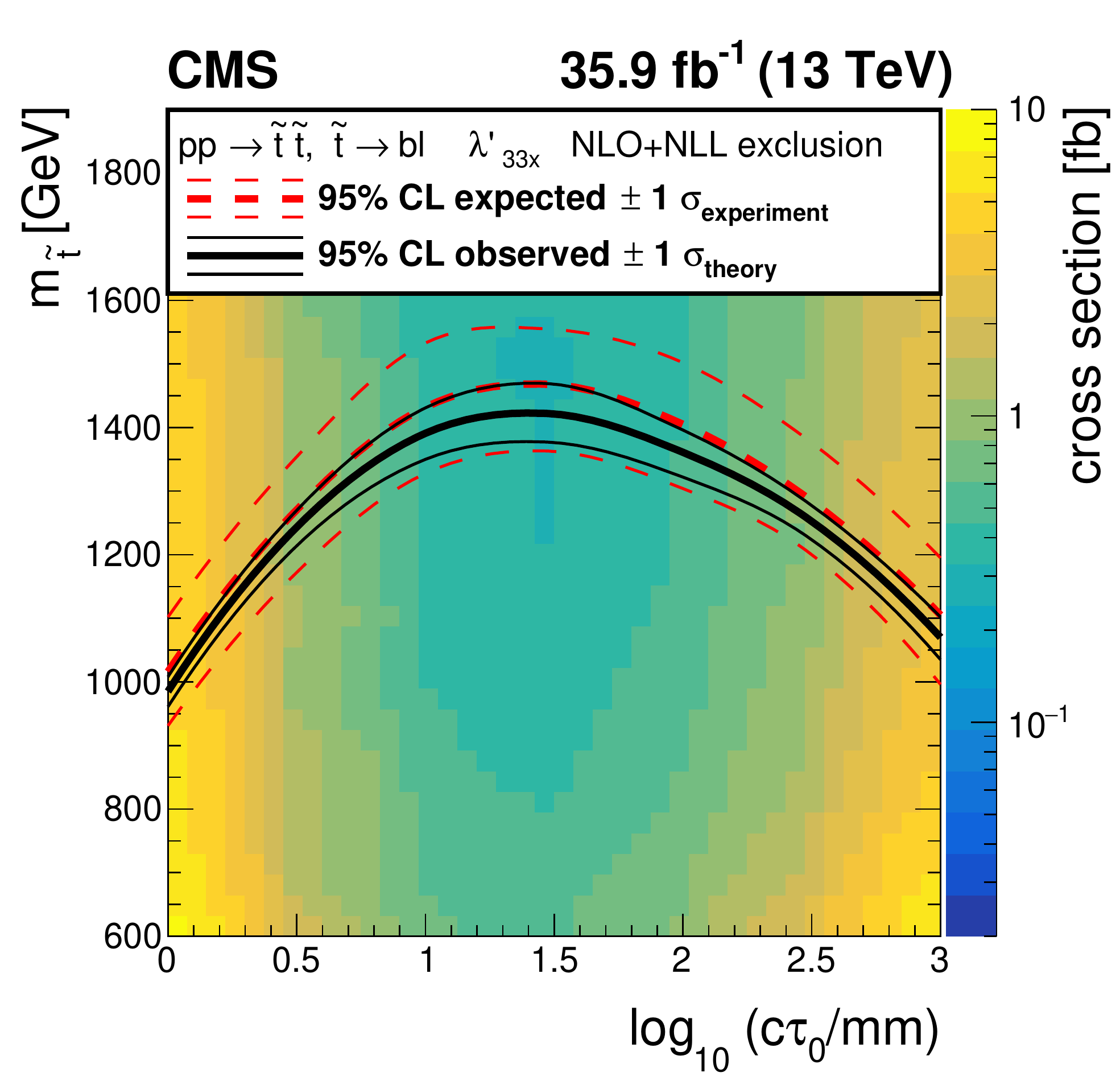}
\caption{\cmsLeft: the expected and observed $95\%$ $\CL$ upper limits on the pair production cross section of the long-lived top squark, assuming a
$100\%$ branching fraction for $\sTop\to\cPqb\ell$ decays. The horizontal lines indicate the NLO+NLL top squark pair production cross sections for $m_{\sTop}=1600\GeV$ and $m_{\sTop}=1000\GeV$, as well as their variations due to the uncertainties in the choices of renormalization scales, factorization scales, and PDF sets. The solid (dashed) lines represent the observed (median expected) limits, the bands show the regions containing $68\%$
of the distributions of the expected limits under the background-only hypothesis.
\cmsRight: the expected and observed $95\%$ limits for the long-lived top squark model in the mass-lifetime plane, assuming a $100\%$ branching fraction
for $\sTop\to \cPqb\ell$ decays, based on the NLO+NLL calculation of the top squark pair production cross section at $\sqrt{s}=13\TeV$. The thick solid black (dashed red) line represents the observed (median expected) limits at $95\%$ $\CLnp$. The thin black lines represent the change in the observed limit
due to the variation of the signal cross sections within their theoretical uncertainties. The thin red lines indicate the region containing $68\%$ of
the distributions of the expected limits under the background-only hypothesis.}
\label{fig: expected_limit_RPV}
\end{figure}

Figure~\ref{fig: expected_limit_RPV} presents the expected and observed upper limits on the pair production cross section
of the long-lived top squark in the RPV $\sTop\to\cPqb\ell$ model, assuming a $100\%$ branching fraction for the top squark to decay to a bottom quark and a charged lepton. The upper limits on the pair production cross section are then
translated into upper limits on the top squark mass for different proper decay lengths, based on an NLO+NLL calculation of the top squark
pair production cross section at $\sqrt{s}=13\TeV$~\cite{Beenakker:1996ch,Kulesza:2008jb,Kulesza:2009kq,Beenakker:2011fu,Borschensky:2014cia} in the limit where all the other SUSY particles are much heavier and decoupled. Top squark masses up to $1350\GeV$ are excluded for proper decay lengths between 7 and $110\mm$. The bounds are currently the most stringent on this model for proper decay lengths between $3\mm$ and 1\unit{m}.

\begin{figure}[tbp]
\centering
\includegraphics[width=0.45\textwidth]{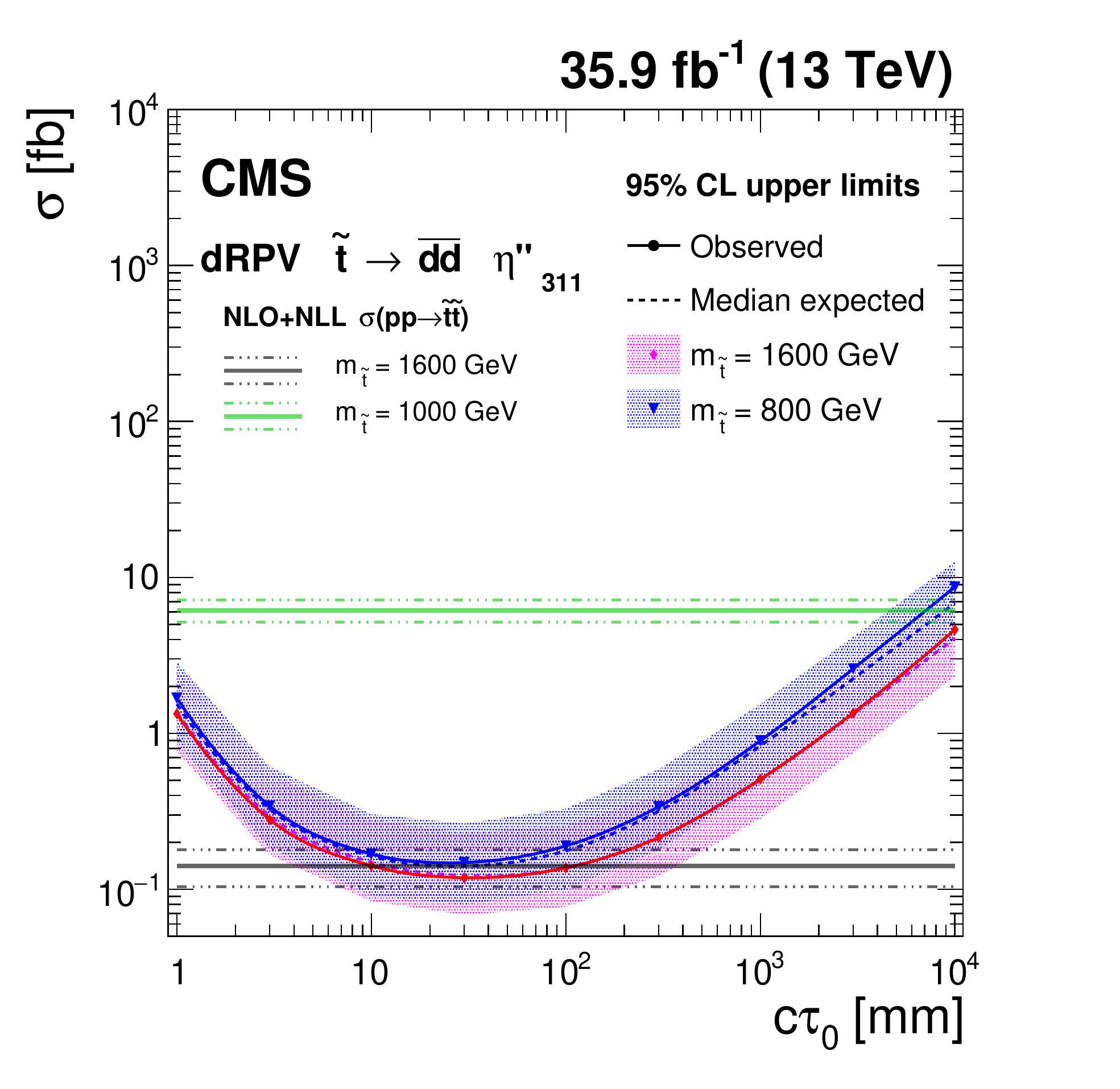}
\includegraphics[width=0.45\textwidth]{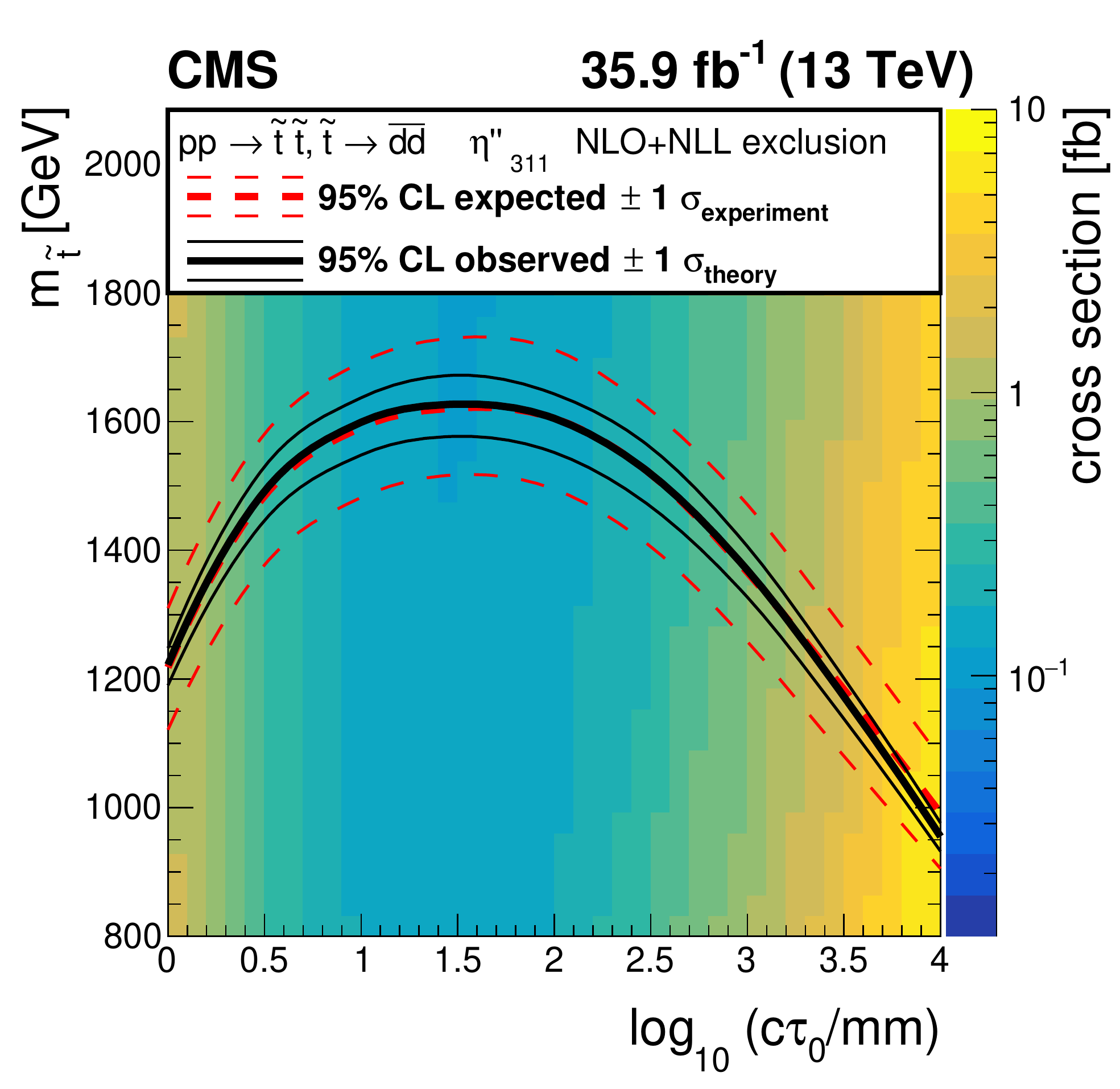}
\caption{\cmsLeft: the expected and observed $95\%$ $\CL$ upper limits on the the pair production cross section of the long-lived top squark, assuming
a $100\%$ branching fraction for $\sTop\to\cPaqd\cPaqd$ decays. The horizontal lines indicate the NLO+NLL top squark pair production cross sections for $m_{\sTop}=1600\GeV$ and $m_{\sTop}=1000\GeV$, as well as their variations due to the uncertainties in the choices of renormalization scales, factorization scales, and PDF sets. The solid (dashed) lines represent the observed (median expected) limits, the bands show the regions containing $68\%$
of the distributions of the expected limits under the background-only hypothesis.
\cmsRight: the expected and observed $95\%$ limits for the long-lived top squark model in the mass-lifetime plane, assuming a $100\%$ branching fraction
for $\sTop\to\cPaqd\cPaqd$ decays, based on an NLO+NLL calculation of the top squark pair production cross section at $\sqrt{s}=13\TeV$. The thick solid black (dashed red) line represents the observed (median expected) limits at $95\%$ $\CLnp$. The thin black lines represent the change in the observed limit
due to the variation of the signal cross sections within their theoretical uncertainties. The thin red lines indicate the region containing $68\%$ of
the distribution of the expected limits under the background-only hypothesis.}
\label{fig: expected_limit_StopTodd}
\end{figure}

Figure~\ref{fig: expected_limit_StopTodd} presents the expected and observed
upper limits on the pair production cross section of the long-lived top squark in the dRPV $\sTop\to\cPaqd\cPaqd$ model,
assuming a $100\%$ branching fraction for the top squark to decay to two down antiquarks. The upper limits on the pair production cross
section are translated into upper limits on the top squark mass for different
proper decay lengths assuming a $100\%$ branching fraction, based on the NLO+NLL calculation of the
top squark pair production cross section at $\sqrt{s}=13\TeV$~\cite{Beenakker:1996ch,Kulesza:2008jb,Kulesza:2009kq,Beenakker:2011fu,Borschensky:2014cia} in the limit where all the other SUSY particles are much heavier and decoupled. Top squark masses up to $1600\GeV$ are excluded for proper decay lengths between
10 and $100\mm$. The bounds are currently the most stringent on this model for
proper decay lengths between $10\mm$ and 10\unit{m}. A comparison on this model with the
existing CMS search for displaced vertices within the beam pipe~\cite{Sirunyan:2018pwn} can be found in Fig.~\ref{fig: expected_limit_comparison} of Appendix~\ref{app:suppMat}.

\section{Summary}

A search for long-lived particles decaying to jets is presented, based on proton-proton collision data collected with the CMS experiment at a
center-of-mass energy of $13\TeV$ in 2016, corresponding to an integrated luminosity of 35.9\fbinv.
The analysis utilizes a dedicated trigger to capture events with displaced-jet signatures, and exploits jet, track, and secondary vertex
information to discriminate displaced-jet candidate events from those produced by the standard model and instrumental backgrounds. The observed yields in data are in agreement with the background predictions.
For a variety of models, we set the best limits to date for long-lived particles with proper decay lengths approximately between $5\mm$ and 10\unit{m}.
Upper limits are set at 95\% confidence level on the pair production cross section of long-lived neutral particles decaying to two jets, for different masses and proper lifetimes, and are as low as 0.2\unit{fb} at high mass ($m_{\mathrm{X}}>1000\GeV$) for proper decay lengths
between 3 and 130\mm. A supersymmetric (SUSY) model with gauge-mediated supersymmetry breaking (GMSB) is also
tested, in which the long-lived gluino can decay to one jet and a lightest SUSY particle. Upper limits are set on the pair production cross section
of the gluino with different masses and proper decay lengths $c\tau_{0}$. Pair-produced long-lived gluinos lighter than $2300\GeV$ are excluded for proper
decay lengths between 20 and $110\mm$. For an $R$-parity violating (RPV) SUSY model, where the long-lived gluino can
decay to top, bottom, and strange antiquarks, pair-produced gluinos lighter than $2400\GeV$
are excluded for decay lengths between 10 and $250\mm$. For a second RPV SUSY model, in which the long-lived top squark can decay to one bottom quark and a charged lepton,
pair-produced long-lived top squarks lighter than $1350\GeV$ are excluded for decay lengths between 7 and $110\mm$. For another RPV SUSY model where
the long-lived top squark decays to two down antiquarks, pair-produced long-lived top
squarks lighter than $1600\GeV$ are excluded for decay lengths between 10 and $110\mm$. These are the most
stringent limits to date on these models.

\begin{acknowledgments}
We congratulate our colleagues in the CERN accelerator departments for the excellent performance of the LHC and thank the technical and administrative staffs at CERN and at other CMS institutes for their contributions to the success of the CMS effort. In addition, we gratefully acknowledge the computing centers and personnel of the Worldwide LHC Computing Grid for delivering so effectively the computing infrastructure essential to our analyses. Finally, we acknowledge the enduring support for the construction and operation of the LHC and the CMS detector provided by the following funding agencies: BMBWF and FWF (Austria); FNRS and FWO (Belgium); CNPq, CAPES, FAPERJ, FAPERGS, and FAPESP (Brazil); MES (Bulgaria); CERN; CAS, MoST, and NSFC (China); COLCIENCIAS (Colombia); MSES and CSF (Croatia); RPF (Cyprus); SENESCYT (Ecuador); MoER, ERC IUT, and ERDF (Estonia); Academy of Finland, MEC, and HIP (Finland); CEA and CNRS/IN2P3 (France); BMBF, DFG, and HGF (Germany); GSRT (Greece); NKFIA (Hungary); DAE and DST (India); IPM (Iran); SFI (Ireland); INFN (Italy); MS and NRF (Republic of Korea); MES (Latvia); LAS (Lithuania); MOE and UM (Malaysia); BUAP, CINVESTAV, CONACYT, LNS, SEP, and UASLP-FAI (Mexico); MOS (Montenegro); MBIE (New Zealand); PAEC (Pakistan); MSHE and NSC (Poland); FCT (Portugal); JINR (Dubna); MON, RosAtom, RAS, RFBR, and NRC KI (Russia); MESTD (Serbia); SEIDI, CPAN, PCTI, and FEDER (Spain); MOSTR (Sri Lanka); Swiss Funding Agencies (Switzerland); MST (Taipei); ThEPCenter, IPST, STAR, and NSTDA (Thailand); TUBITAK and TAEK (Turkey); NASU and SFFR (Ukraine); STFC (United Kingdom); Department of Energy and National Science Foundation (USA).

\hyphenation{Rachada-pisek} Individuals have received support from the Marie-Curie program and the European Research Council and Horizon 2020 Grant, Contract No. 675440 (European Union); the Leventis Foundation; the A. P. Sloan Foundation; the Alexander von Humboldt Foundation; the Belgian Federal Science Policy Office; the Fonds pour la Formation \`a la Recherche dans l'Industrie et dans l'Agriculture (FRIA-Belgium); the Agentschap voor Innovatie door Wetenschap en Technologie (IWT-Belgium); the F.R.S.-FNRS and FWO (Belgium) under the ``Excellence of Science - EOS" - be.h project No. 30820817; the Ministry of Education, Youth and Sports (MEYS) of the Czech Republic; the Lend\"ulet (``Momentum") Programme and the J\'anos Bolyai Research Scholarship of the Hungarian Academy of Sciences, the New National Excellence Program \'UNKP, the NKFIA Research Grants No. 123842, No. 123959, No. 124845, No. 124850 and No. 125105 (Hungary); the Council of Science and Industrial Research, India; the HOMING PLUS program of the Foundation for Polish Science, cofinanced from European Union, Regional Development Fund, the Mobility Plus program of the Ministry of Science and Higher Education, the National Science Center (Poland), Contracts Harmonia No. 2014/14/M/ST2/00428, Opus No. 2014/13/B/ST2/02543, No. 2014/15/B/ST2/03998, and No. 2015/19/B/ST2/02861, Sonata-bis No. 2012/07/E/ST2/01406; the National Priorities Research Program by Qatar National Research Fund; the Programa Estatal de Fomento de la Investigaci{\'o}n Cient{\'i}fica y T{\'e}cnica de Excelencia Mar\'{\i}a de Maeztu, Grant No. MDM-2015-0509 and the Programa Severo Ochoa del Principado de Asturias; the Thalis and Aristeia programs cofinanced by EU-ESF and the Greek NSRF; the Rachadapisek Sompot Fund for Postdoctoral Fellowship, Chulalongkorn University and the Chulalongkorn Academic into Its 2nd Century Project Advancement Project (Thailand); the Welch Foundation, Contract No. C-1845; and the Weston Havens Foundation (USA).
\end{acknowledgments}

\bibliography{auto_generated}
\clearpage
\appendix
\section{Supplemental information}\label{app:suppMat}
\begin{figure*}[htb]
\centering
\includegraphics[width=0.45\textwidth]{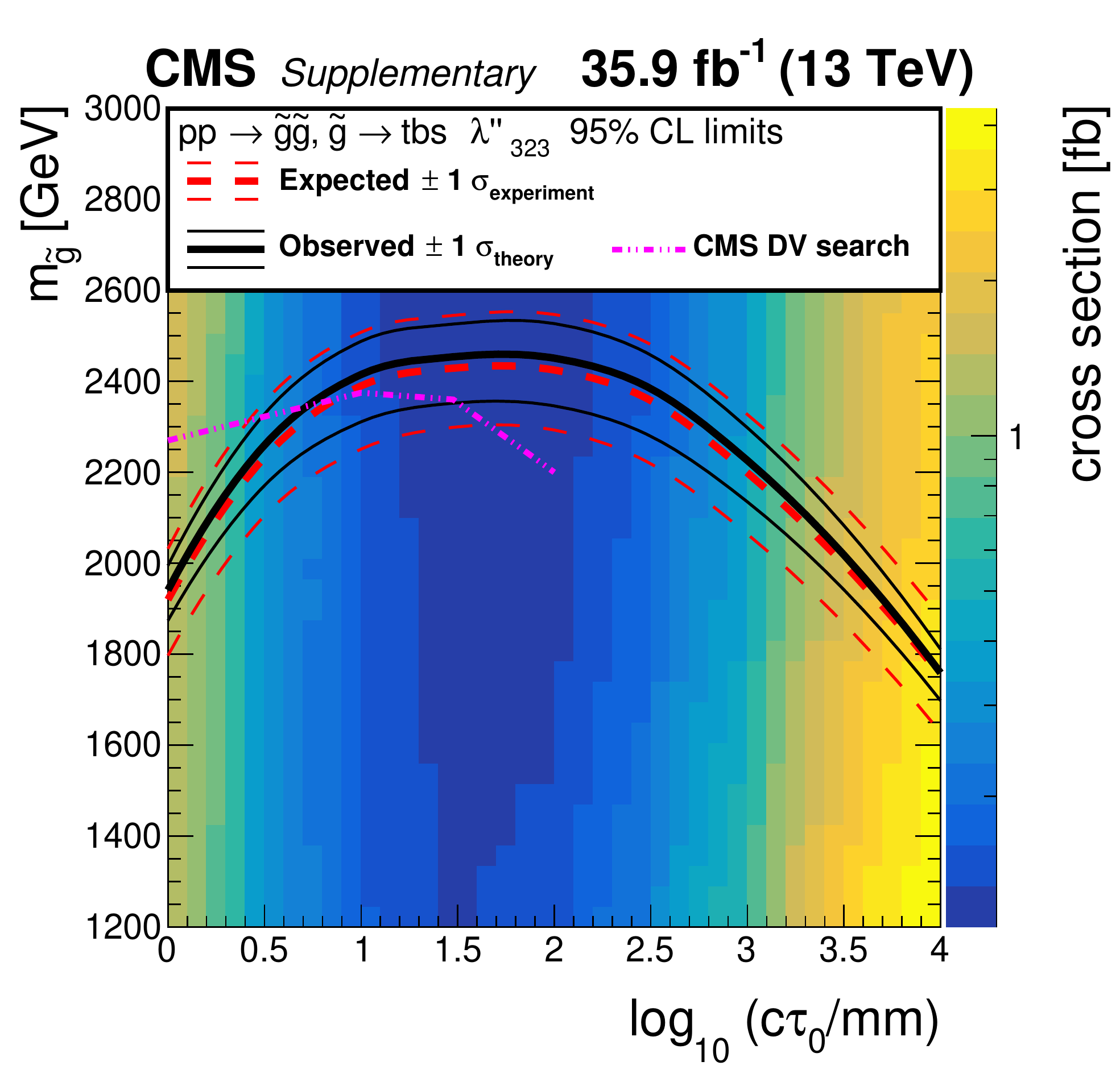}
\includegraphics[width=0.45\textwidth]{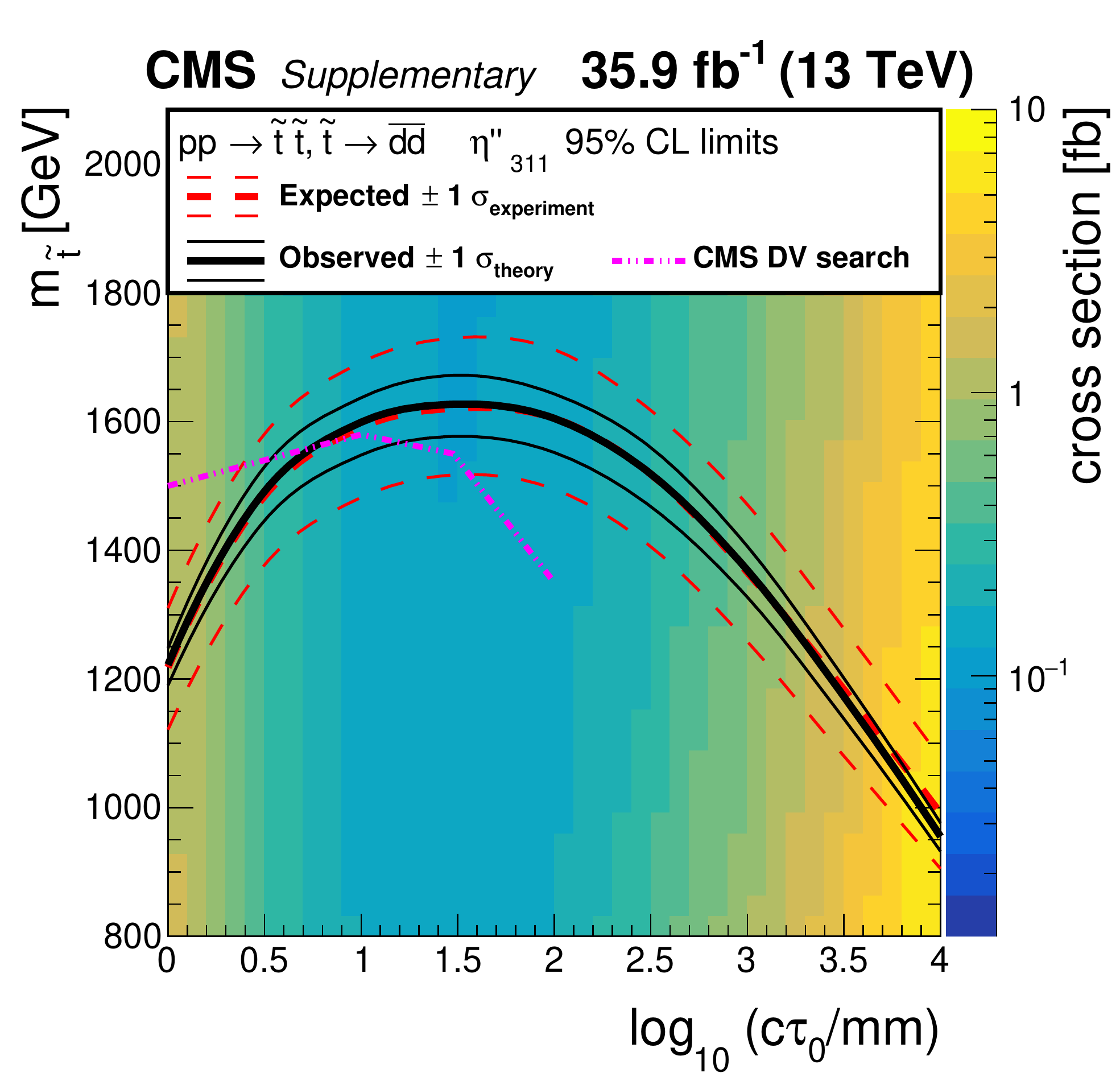}
\caption{Comparison with search for displaced vertices in multijet events at $\sqrt{s}=13\TeV$ with the CMS detector~\cite{Sirunyan:2018pwn} (referred to as the CMS DV search) for $\sGlu\to\cPqt\cPqb\cPqs$ (left) and $\sTop\to\cPaqd\cPaqd$ (right) models. The CMS DV search looks for a pair of displaced vertices within the beam pipe. The observed limits obtained by the CMS DV search (purple curves) are overlaid with the limits obtained by the search presented in this paper
in the mass-lifetime plane, and are good complements for proper decay length $c\tau_{0}<10\mm$ in these two signal models.}
\label{fig: expected_limit_comparison}
\end{figure*}

Tables~\ref{tab: eff_JetJet}--\ref{tab: eff_tdd} summarize the signal efficiencies for representative signal points in jet-jet, $\sGlu\to\Glu\sGra$, $\sGlu\to\cPqt\cPqb\cPqs$,
$\sTop\to\cPqb\ell$, and $\sTop\to\cPaqd\cPaqd$ models. Figure~\ref{fig: expected_limit_comparison} shows the comparison with the search
for displaced vertices in multijet events at $\sqrt{s}=13\TeV$ with the CMS detector~\cite{Sirunyan:2018pwn}, for $\sGlu\to\cPqt\cPqb\cPqs$ and $\sTop\to\cPaqd\cPaqd$ models.

\begin{table*}[hbt]
  \topcaption{ Signal efficiencies (in \%) for the jet-jet model at different proper decay
lengths $c\tau_{0}$ and different masses $m_{\mathrm{X}}$. Selection requirements are cumulative from
    the first row to the last. Uncertainties are statistical only. \label{tab:cutflow_XXTo4J}}
\label{tab: eff_JetJet}
\centering

\begin{scotch}{lccccc}
\multirow{2}*{Efficiency ($\%$)}&\multirow{2}*{$m_{\mathrm{X}}$ ($\GeV$)} & \multicolumn{4}{c}{c$\tau_{0}$}\\
 & & $1\mm$ & $10\mm$ & $100\mm$ & $1000\mm$ \\\hline
Trigger & \multirow{3}*{1000} & $20.32\pm0.45$ & $82.96\pm0.91$ & 64.58 $\pm$ 0.80 & 12.94 $\pm$ 0.36 \\
Preselection & & 17.99 $\pm$ 0.42 & 80.54 $\pm$ 0.90 & 61.40 $\pm$ 0.78 & 11.29 $\pm$ 0.34 \\
Final selection & & 9.69 $\pm$ 0.31  & 57.23 $\pm$ 0.76 & 44.86 $\pm$ 0.67 & 7.79 $\pm$ 0.28 \\
\multicolumn{6}{c}{}\\
Trigger & \multirow{3}*{300} & 18.86 $\pm$ 0.43 & 69.00 $\pm$ 0.83 & 42.44 $\pm$ 0.65 & 6.27 $\pm$ 0.25 \\
Preselection & & 14.22 $\pm$ 0.37 & 60.94 $\pm$ 0.78 & 36.53 $\pm$ 0.60 & 4.83 $\pm$ 0.22 \\
Final selection & & 6.98  $\pm$ 0.26 & 39.51 $\pm$ 0.63 & 22.0 $\pm$ 0.47 & 2.82 $\pm$ 0.17 \\
\multicolumn{6}{c}{}\\
Trigger & \multirow{3}*{100} & 3.10 $\pm$ 0.10 & 10.3 $\pm$ 0.18 & 4.91 $\pm$ 0.13 & 0.50 $\pm$ 0.04 \\
Preselection & & 2.13 $\pm$ 0.08 & 7.91 $\pm$ 0.16 & 3.48 $\pm$ 0.11 & 0.34 $\pm$ 0.03 \\
Final selection & & 0.92 $\pm$ 0.06 & 4.41 $\pm$ 0.12 & 1.64 $\pm$ 0.07 & 0.17 $\pm$ 0.02 \\
\end{scotch}
\end{table*}

\begin{table*}[hbt]
  \topcaption{ Signal efficiencies (in \%) for pair produced long-lived gluinos decaying to a gluon
and a gravitino at different proper decay
lengths $c\tau_{0}$ and different gluino masses $m_{\sGlu}$. Selection requirements are cumulative from
    the first row to the last. Uncertainties are statistical only.\label{tab:cutflow_GMSB}}
\label{tab: eff_ggG}
\centering
\begin{scotch}{lccccc}
\multirow{2}*{Efficiency ($\%$)}&\multirow{2}*{$m_{\sGlu}$ ($\GeV$)} & \multicolumn{4}{c}{c$\tau_{0}$}\\
 & & $1\mm$ & $10\mm$ & $100\mm$ & $1000\mm$ \\\hline
Trigger & \multirow{3}*{2400} & 6.58 $\pm$ 0.12 & 62.49 $\pm$ 0.35 & 75.45 $\pm$ 0.39 & 27.55 $\pm$ 0.25 \\
Preselection & & 4.80 $\pm$ 0.10 & 57.62 $\pm$ 0.34 & 68.36 $\pm$ 0.37 & 22.58 $\pm$ 0.21 \\
Final selection & & 2.02 $\pm$ 0.06 & 31.73 $\pm$ 0.25 & 43.45 $\pm$ 0.29 & 14.18 $\pm$ 0.17  \\
\multicolumn{6}{c}{}\\
Trigger & \multirow{3}*{1800} & 7.43 $\pm$ 0.12 & 61.48 $\pm$ 0.35 & 70.12 $\pm$ 0.37 & 22.31 $\pm$ 0.21 \\
Preselection & & 5.56 $\pm$ 0.11 & 56.05 $\pm$ 0.33 & 62.05 $\pm$ 0.35 & 18.05 $\pm$ 0.19 \\
Final selection & & 2.19 $\pm$ 0.07 & 29.91 $\pm$ 0.24 & 37.73 $\pm$ 0.27 & 10.88 $\pm$ 0.15 \\
\multicolumn{6}{c}{}\\
Trigger & \multirow{3}*{1000}  & 7.42 $\pm$ 0.13 & 55.92 $\pm$ 0.34 & 57.58 $\pm$ 0.34 & 13.71 $\pm$ 0.16 \\
Preselection & & 5.47 $\pm$ 0.11 & 48.55 $\pm$ 0.31 & 47.13 $\pm$ 0.31 & $10.52\pm0.15$ \\
Final selection & & 1.96 $\pm$ 0.06 & 23.48 $\pm$ 0.22 & 25.78 $\pm$ 0.23 & 5.52 $\pm$ 0.11 \\
\end{scotch}
\end{table*}

\begin{table*}[hbt]
 \topcaption{ Signal efficiencies (in \%) for pair produced long-lived gluinos decaying to top, bottom and strange antiquarks at different proper decay lengths $c\tau_{0}$ and different gluino masses $m_{\sGlu}$. Selection requirements are cumulative from the first row to the last. Uncertainties
are statistical only.}
\label{tab: eff_gtbs}
\centering
\begin{scotch}{lccccc}
\multirow{2}*{Efficiency ($\%$)}&\multirow{2}*{$m_{\sGlu}$ ($\GeV$)} & \multicolumn{4}{c}{c$\tau_{0}$}\\
 & & $1\mm$ & $10\mm$ & $100\mm$ & $1000\mm$ \\\hline
Trigger & \multirow{3}*{2400} & 12.53 $\pm$ 0.41 & 80.30 $\pm$ 4.90 & 85.00 $\pm$ 1.10 & 33.83 $\pm$ 0.41 \\
Preselection & & 10.48 $\pm$ 0.37 & 79.70 $\pm$ 4.88 & 84.57 $\pm$ 1.10 & 32.01 $\pm$ 0.40 \\
Final selection & & 4.42 $\pm$ 0.24 & 51.04 $\pm$ 3.90 & 60.35 $\pm$ 0.93 & 21.55 $\pm$ 0.33  \\
\multicolumn{6}{c}{}\\
Trigger & \multirow{3}*{1800} & 14.95 $\pm$ 0.28 & 78.94 $\pm$ 0.64 & 82.93 $\pm$ 0.65 & 28.81 $\pm$ 0.38 \\
Preselection & & 10.48 $\pm$ 0.37 & 79.70 $\pm$ 4.88 & 84.57 $\pm$ 1.10 & 32.01 $\pm$ 0.40 \\
Final selection & & 4.42 $\pm$ 0.24 & 51.04 $\pm$ 3.90 & 60.35 $\pm$ 0.93 & 21.55 $\pm$ 0.33  \\
\multicolumn{6}{c}{}\\
Trigger & \multirow{3}*{1200} & 18.30 $\pm$ 0.30 & 78.32 $\pm$ 0.63 & 77.75 $\pm$ 0.63 & 23.39 $\pm$ 0.34 \\
Preselection & & 15.21 $\pm$ 0.28 & 76.92 $\pm$ 0.62 & 76.94 $\pm$ 0.63 & $21.45\pm0.33$ \\
Final selection & & 5.21 $\pm$ 0.16 & 43.01 $\pm$ 0.47 & 48.40 $\pm$ 0.50 & 12.03 $\pm$ 0.24 \\
\end{scotch}
\end{table*}

\begin{table*}[hbt]
  \topcaption{ Signal efficiencies (in \%) for pair produced long-lived top squarks decaying to a bottom quark
and a lepton at different proper decay
lengths $c\tau_{0}$ and different top squark masses $m_{\sTop}$. Selection requirements are cumulative from
    the first row to the last. Uncertainties are statistical only.}
\label{tab: eff_tbl}
\centering
\begin{scotch}{lccccc}
\multirow{2}*{Efficiency ($\%$)}&\multirow{2}*{$m_{\sTop}$ ($\GeV$)} & \multicolumn{4}{c}{c$\tau_{0}$}\\
 & & $1\mm$ & $10\mm$ & $100\mm$ & $1000\mm$ \\\hline
Trigger & \multirow{3}*{1500} & 6.44 $\pm$ 0.19 & 48.12 $\pm$ 0.52 & 46.33 $\pm$ 0.53 & 12.02 $\pm$ 0.26 \\
Preselection & & 3.79 $\pm$ 0.14 & 37.11 $\pm$ 0.45 & 32.39 $\pm$ 0.45 & 6.24 $\pm$ 0.19 \\
Final selection & & 1.57 $\pm$ 0.09 & 20.53 $\pm$ 0.34 & 17.47 $\pm$ 0.33 & 3.02 $\pm$ 0.13  \\
\multicolumn{6}{c}{}\\
Trigger & \multirow{3}*{1200} & 6.68 $\pm$ 0.09 & 46.25 $\pm$ 0.23 & 42.98 $\pm$ 0.21 & 8.71$\pm$ 0.10 \\
Preselection & & 4.34 $\pm$ 0.07 & 38.28 $\pm$ 0.21 & 33.62 $\pm$ 0.19 & 5.83 $\pm$ 0.08 \\
Final selection & & 1.55 $\pm$ 0.04 & 18.41 $\pm$ 0.14 & 16.54 $\pm$ 0.14 & 2.63 $\pm$ 0.05 \\
\multicolumn{6}{c}{}\\
Trigger & \multirow{3}*{600} & 6.99 $\pm$ 0.08 & 41.12 $\pm$ 0.21 & 32.65 $\pm$ 0.19 & 5.28 $\pm$ 0.08 \\
Preselection & & 3.53 $\pm$ 0.06 & 29.69 $\pm$ 0.18 & 22.34 $\pm$ 0.16 & 3.06 $\pm$ 0.06 \\
Final selection & & 0.88 $\pm$ 0.03 & 11.63 $\pm$ 0.11 & 8.71 $\pm$ 0.10 & 1.10 $\pm$ 0.03  \\
\end{scotch}
\end{table*}

\begin{table*}[hbt]
  \topcaption{ Signal efficiencies (in \%) for pair produced long-lived top squarks decaying to two down antiquarks at different proper decay
lengths $c\tau_{0}$ and different top squark masses $m_{\sTop}$. Selection requirements are cumulative from
    the first row to the last. Uncertainties are statistical only.\label{tab:cutflow_Stopdd}}
\label{tab: eff_tdd}
\centering
\begin{scotch}{lccccc}
\multirow{2}*{Efficiency ($\%$)}&\multirow{2}*{$m_{\sTop}$ ($\GeV$)} & \multicolumn{4}{c}{c$\tau_{0}$}\\
 & & $1\mm$ & $10\mm$ & $100\mm$ & $1000\mm$ \\\hline
Trigger & \multirow{3}*{1600} & 12.05 $\pm$ 0.25 & 74.66 $\pm$ 0.62 & 77.19 $\pm$ 0.98 & 24.03 $\pm$ 0.54 \\
Preselection & & 10.27 $\pm$ 0.23 & 72.70 $\pm$ 0.61 & 75.01 $\pm$ 0.97 & 21.43 $\pm$ 0.51 \\
Final selection & & 5.36 $\pm$ 0.16 & 48.75 $\pm$ 0.50 & 53.59 $\pm$ 0.81 & 14.65 $\pm$ 0.42  \\
\multicolumn{6}{c}{}\\
Trigger & \multirow{3}*{1200} & 12.31 $\pm$ 0.25 & 73.74 $\pm$ 0.61 & 73.92 $\pm$ 0.61 & $20.26\pm0.48$ \\
Preselection & & 10.46 $\pm$ 0.23 & 71.55 $\pm$ 0.60 & 71.36 $\pm$ 0.60 & 18.04 $\pm$ 0.45 \\
Final selection & & 5.13 $\pm$ 0.16 & 48.04 $\pm$ 0.49 & 49.44 $\pm$ 0.50 & 12.27 $\pm$ 0.37 \\
\multicolumn{6}{c}{}\\
Trigger & \multirow{3}*{600} & 12.02 $\pm$ 0.37 & 71.75 $\pm$ 0.89 & 67.03 $\pm$ 0.92 & 16.27 $\pm$ 0.29 \\
Preselection &  & 9.97 $\pm$ 0.33 & 69.08 $\pm$ 0.88 & 63.60 $\pm$ 0.89 & 14.38 $\pm$ 0.26 \\
Final selection &  & 4.90 $\pm$ 0.23 & 45.68 $\pm$ 0.71 & 42.61 $\pm$ 0.72 & 9.08 $\pm$ 0.21  \\
\end{scotch}
\end{table*}

\cleardoublepage \section{The CMS Collaboration \label{app:collab}}\begin{sloppypar}\hyphenpenalty=5000\widowpenalty=500\clubpenalty=5000\vskip\cmsinstskip
\textbf{Yerevan Physics Institute, Yerevan, Armenia}\\*[0pt]
A.M.~Sirunyan, A.~Tumasyan
\vskip\cmsinstskip
\textbf{Institut f\"{u}r Hochenergiephysik, Wien, Austria}\\*[0pt]
W.~Adam, F.~Ambrogi, E.~Asilar, T.~Bergauer, J.~Brandstetter, M.~Dragicevic, J.~Er\"{o}, A.~Escalante~Del~Valle, M.~Flechl, R.~Fr\"{u}hwirth\cmsAuthorMark{1}, V.M.~Ghete, J.~Hrubec, M.~Jeitler\cmsAuthorMark{1}, N.~Krammer, I.~Kr\"{a}tschmer, D.~Liko, T.~Madlener, I.~Mikulec, N.~Rad, H.~Rohringer, J.~Schieck\cmsAuthorMark{1}, R.~Sch\"{o}fbeck, M.~Spanring, D.~Spitzbart, W.~Waltenberger, J.~Wittmann, C.-E.~Wulz\cmsAuthorMark{1}, M.~Zarucki
% ...
\vskip\cmsinstskip
% ...
1:  Also at Vienna University of Technology, Vienna, Austria\\
% ...
\end{sloppypar}
\end{document}